\documentclass[a4paper]{elsart5p}
\usepackage{amsmath}
\usepackage{txfonts}
\usepackage[utf8x]{inputenc}
\usepackage{graphicx}
\usepackage[numbers]{natbib}
\usepackage[protrusion=true,expansion=true]{microtype}
\def\includeplaatje[#1]#2{\includegraphics[#1]{#2.pdf}}%
\journal{Astroparticle Physics}

\def\e#1{\cdot10^{#1}}
\def\E#1{\mathrm{e}^{#1}}

\def\pardiff#1#2{\frac{\partial #1}{\partial #2}}
\def\max{\mathrm{max}}
\def\min{\mathrm{min}}
\def\corsika{\textsc{corsika}}
\def\coast{\textsc{coast}}

\def\lepton{\mathrm{e}}
\def\photon{\gamma}
\def\pion{\pi}
\def\muon{\mu}
\def\neutrino{\nu}
\let\upGamma\Gamma
\def\unit#1{\ifmmode\text{~#1}\else~#1\fi}
\def\gcm{\ifmmode\textrm{g}/\textrm{cm}^2\else$\textrm{g}/\textrm{cm}^2$\fi}
\let\figurewidth\columnwidth
\begin{document}
\begin{frontmatter}

\title{Universality of electron-positron distributions in extensive air showers}
\author[nijm]{S.~Lafebre\corauthref{cor}},
\corauth[cor]{Corresponding author. Now at: Center for Particle Astrophysics, Pennsylvania State University, University Park, PA 16802, United States. Tel. +1\,814\,865\,0979; fax: +1\,814\,863\,3297}
\ead{s.lafebre@astro.ru.nl}
\author[fzk]{R.~Engel},
\author[nijm,astron]{H.~Falcke},
\author[nijm]{J.~H\"orandel},
\author[fzk]{T.~Huege},
\author[nijm]{J.~Kuijpers},
\author[fzk]{R.~Ulrich}

\address[nijm]
    {Department of Astrophysics, IMAPP, Radboud University, P.O.~Box 9010, 6500GL Nijmegen, The Netherlands}
\address[fzk]
    {Institut f\"ur Kernphysik, Forschungszentrum Karlsruhe, P.O.~Box 3640, 76021 Karlsruhe, Germany}
\address[astron]
    {Radio Observatory, Astron, Dwingeloo, P.O.~Box 2, 7990AA Dwingeloo, The Netherlands}

\date{Received $\langle$date$\rangle$ / Accepted $\langle$date$\rangle$}

\begin{abstract}
Using a large set of simulated extensive air showers, we investigate universality features of electron and positron distributions in very-high-energy cosmic-ray air showers. Most particle distributions depend only on the depth of the shower maximum and the number of particles in the cascade at this depth. We provide multi-dimensional parameterizations for the electron-positron distributions in terms of particle energy, vertical and horizontal momentum angle, lateral distance, and time distribution of the shower front. These parameterizations can be used to obtain realistic electron-positron distributions in extensive air showers for data analysis and simulations of Cherenkov radiation, fluorescence signal, and radio emission.
\end{abstract}

\begin{keyword}
Ultra-high-energy cosmic rays; Extensive air showers; Electron angular distributions; Electron energy spectra; Electron lateral distributions; Shower front structure
\PACS 96.50.S- \sep 96.50.sd
\end{keyword}
\end{frontmatter}

%
%
\section{Introduction}
%
%
One of the greatest mysteries in particle astrophysics is the nature and origin of the highest-energy cosmic rays above $10^{17}$\unit{eV}. The study of extensive air showers produced in our atmosphere by these particles is the primary means of obtaining information about high-energy cosmic rays. Many techniques to observe these air showers, including the detection of atmospheric fluorescence and Cherenkov light~\cite{1985:Baltrusaitis} and radio signal emission~\cite{2005:Falcke}, depend on the knowledge of the distribution of charged particles in air showers. Primarily, the distributions of electrons and positrons as most abundant charged particles are of importance. Theoretical predictions of the main production and energy loss processes in electromagnetic showers have been available for a long time \citep{1941:Rossi,1965:Nishimura}. Modern Monte Carlo techniques greatly enhance the accuracy of these estimates and allow us to calculate the electron-positron distributions not only in electromagnetic showers but also showers initiated by hadrons.

In this work, we use simulations to investigate electron-positron distributions in extensive air showers and their dependence on energy, species, and zenith angle of the primary particle and on the evolution stage of the shower. Previous studies have shown that many distributions depend only on two parameters: the number of particles in the extensive air shower and the longitudinal position in the shower evolution where this maximum occurs \citep{1982:Hillas:1,2005:Giller,2005:Giller:proc,2006:Gora,2005:Chou,2006:Nerling,2007:Billoir,2008:Schmidt}. This concept, which is referred to as \emph{universality}, allows us to develop parameterizations of the electron-positron distributions as a function of relevant quantities such as energy, lateral distance, and momentum angles, in terms of only a few parameters.

%
%
\section{Method}
        \label{sec:setup}
%
%
Electron and positron distributions in the atmosphere were studied through detailed Monte Carlo simulations. Unless specified otherwise, extensive air shower simulations were performed according to the specifications below.

All simulations were carried out using the \corsika\ code, version~6.5~\cite{corsika:Heck}. We used the \textsc{qgsjet}-II-03 model~\cite{2006:Ostapchenko:2,2006:Ostapchenko:1} to describe high-energy interactions and the \textsc{urqmd}~1.3.1 code~\cite{1998:Bass,1999:Bleicher} at lower energies. Electromagnetic interactions were treated by the \textsc{egs4} code~\cite{1985:Nelson}. We applied a low energy cutoff of $151$\unit{keV} and level~$10^{-6}$ optimum thinning~\cite{2001:Auger,2001:Risse}. The U.S.~Standard Atmosphere~\cite{1976:USatm,corsika:Knapp} was used as atmospheric model. It should be noted that, because simulations for our analysis were performed using only a single nuclear interaction model, the shape of the distributions presented may change somewhat when different models such as \textsc{sibyll} or \textsc{qgsjet}-I are employed. On the other hand, the $e^\pm$ distributions in proton and iron showers exhibit very good universality. Hence, the overall behaviour of the distributions should not change significantly.

The standard output of \corsika\ is a list of momenta, position coordinates, and arrival times of those particles that cross a horizontal plane representing the ground detector. This output format is not ideally suited for universality studies. First of all, particle distributions need to be calculated at many depth layers for each individual shower. Secondly, considering inclined showers, different core distances in the horizontal detector plane correspond to different shower development stages.

A multi-purpose interface called \emph{\coast} (Corsika Data Access Tools) has been developed for accessing the data of individual particles tracked in \corsika~\citep{2007:Ulrich}. For each track segment of a particle simulated in \corsika, a \coast\ interface function is called with the particle properties at the start and end of the propagation step. In addition, all standard \corsika\ output information is passed to the \coast\ interface. This allows one to directly access the overall information of the simulated showers (e.g.\ energy, direction of incidence, depth of first interaction) as well as details on all individual track segments of the simulated shower particles.

The \coast\ interface was used in this work to produce histograms of different particle distributions. Planes perpendicular to the shower axis were defined and particles were filled in the corresponding histograms if their track traversed one of these planes. The energy, momentum, time, and position of a particle crossing one of the planes was calculated by interpolation from the start and end points of the track segment. In total, 50~planes at equidistant levels in slant depth~$X$ between the point of first interaction and sea level ($X\simeq1036$\unit\gcm\ for vertical showers) were used for histogramming, whereas the depth of a plane was measured along the shower axis. Note that these planes are, in general, not horizontal and cover different atmospheric densities. In our universality studies below, we will use only the densities at the intersection points of the planes with the shower axis.

At each of the~50 planes, two three-dimensional histograms were filled for electrons and positrons respectively. The first histogram contains logarithmically binned distributions of the arrival time, lateral distance from the shower axis, and the kinetic energy of the particles. The second histogram contains the angle between the momentum vector and the shower axis, the angle of the momentum vector projected into the plane with respect to the outward direction in the plane, and the kinetic energy of the particles.

Showers were simulated for protons, photons, and iron nuclei at primary energies of $10^{17}$, $10^{18}$, $10^{19}$, and $10^{20}$\unit{eV}. For each combination of primary particle and energy, showers with zenith angles of $0$, $30$, $45$, and~$60$º were calculated. Non-vertical showers were injected from the north, northeast, east, southeast, and south to accommodate deviations due to the geomagnetic field. The preshower effect \citep{1966:Erber,1981:McBreen} was excluded: for photon primaries at energies over $10^{19}$\unit{eV}, it would result in the simulation of several lower-energy primaries, the particle distributions for which are already included. Each parameter set was repeated 20~times, amounting to a total of 3840~simulated showers. The showers were produced with a parallelized \corsika\ version~\cite{2007:Lafebre} on a cluster of $24$~nodes. Access to this library may be obtained through the authors.

As a reference set, averaged distributions at the shower maxima of 20~vertical air showers initiated by $10^{18}$\unit{eV} protons are used. This set is compared to averaged distributions of other parameters, only one of which is changed at a time. If not explicitly stated, all distributions in this work refer to the sum of electrons and positrons. In particular, when the term `particles' is used, the sum of electrons and positrons is meant.

%
%
\section{Longitudinal description}
%
%
There are several ways to describe the longitudinal evolution of an air shower.

\emph{Slant depth}~$X$ measures the amount of matter an air shower has traversed in the atmosphere, in~\gcm.

\emph{Relative evolution stage} is defined here in terms of the depth relative to the slant depth~$X_\max$, where the number of particles in the air shower reaches its maximum
\begin{equation}\label{eq:rdepthdef}
t\equiv\frac{X-X_\max}{X_0},
\end{equation}
with $X_0\simeq36.7$\unit\gcm\ being the radiation length of electrons in air.
Because the shower maximum always lies at $t=0$, describing multiple showers in terms of this quantity rather than~$X$ is expected to lead to a higher degree of universality.

\emph{Shower age} is defined here so that $s=0$ at the top of the atmosphere, $s=1$ at the shower maximum, and $s=3$ at infinite depth
\begin{equation}\label{eq:agedef}
s\equiv\frac{3X}{X+2X_\max}=\frac{t+X_\max/X_0}{t/3+X_\max/X_0}.
\end{equation}
The concept of shower age arises naturally from cascade theory in purely electromagnetic showers~\citep{1941:Rossi,2008:Lipari}. For example, the electron energy distribution is a function of shower age. Eq.~(\ref{eq:agedef}) is, however, only a simple, frequently used phenomenological approximation to the shower age parameter defined in cascade theory. It has the advantage that it can also be applied to showers with a significant hadronic component. Alternatively, shower age could be defined phenomenologically such that $s=0$ corresponds to the depth of the first interaction. Since there is no practical way of observing the depth of the first interaction in air shower measurements this variant is not considered in our analysis.

To determine which description yields the highest degree of universality, electron energy distributions of a sample of 180 showers of various primary energies and initiated by different primaries were compared. Statistical deviations from the average distribution were obtained at fixed relative evolution stages~$t$ and at each individual shower's corresponding value of $X$ and~$s$ according to~\eqref{eq:rdepthdef} and~\eqref{eq:agedef}.

\begin{figure}
    \includeplaatje[width=\figurewidth]{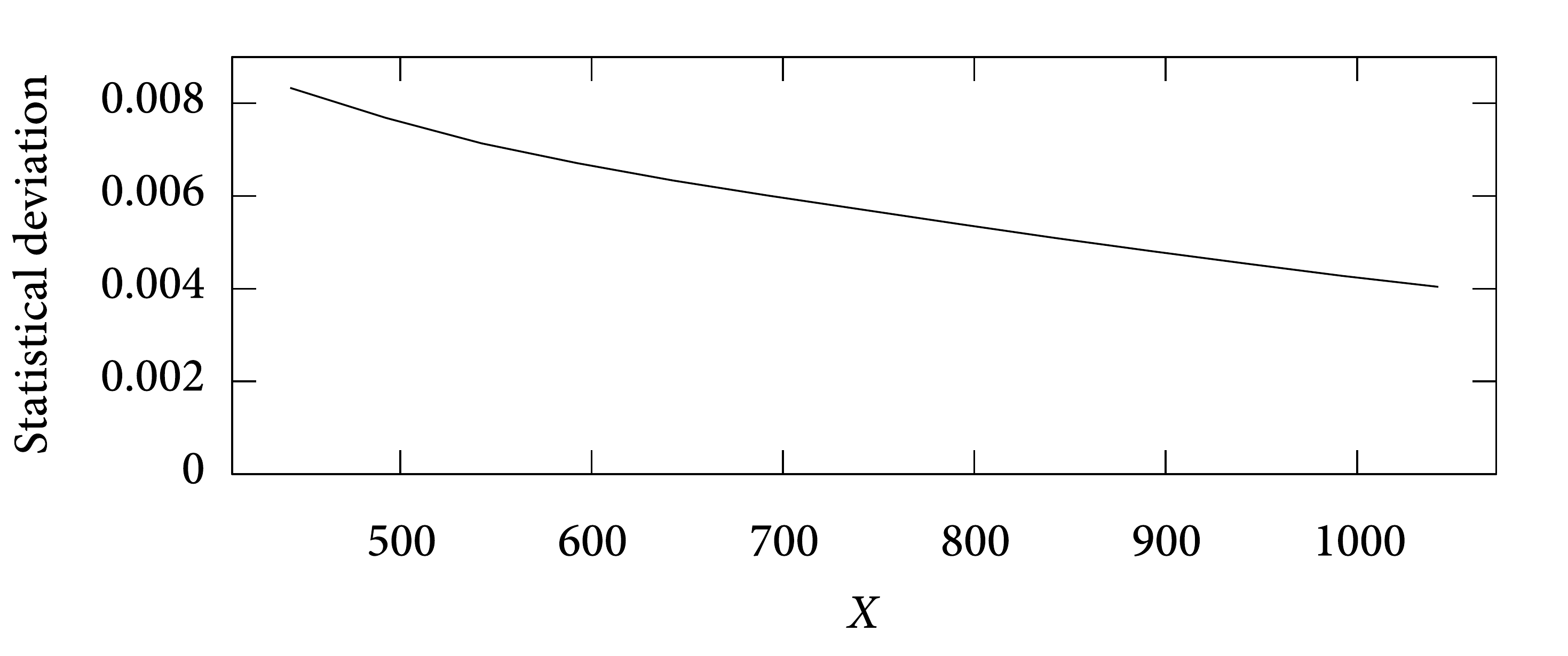}\\[-.02\figurewidth]
    \includeplaatje[width=\figurewidth]{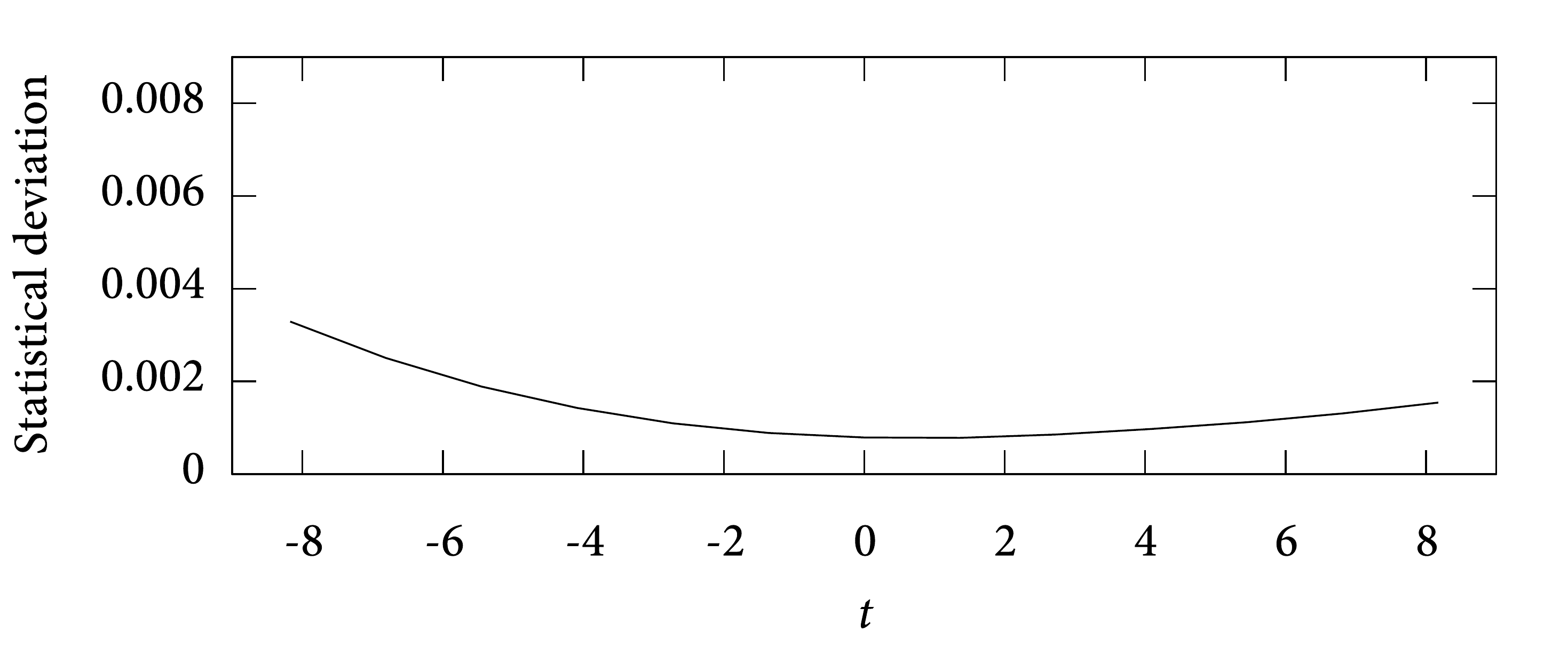}\\[-.02\figurewidth]
    \includeplaatje[width=\figurewidth]{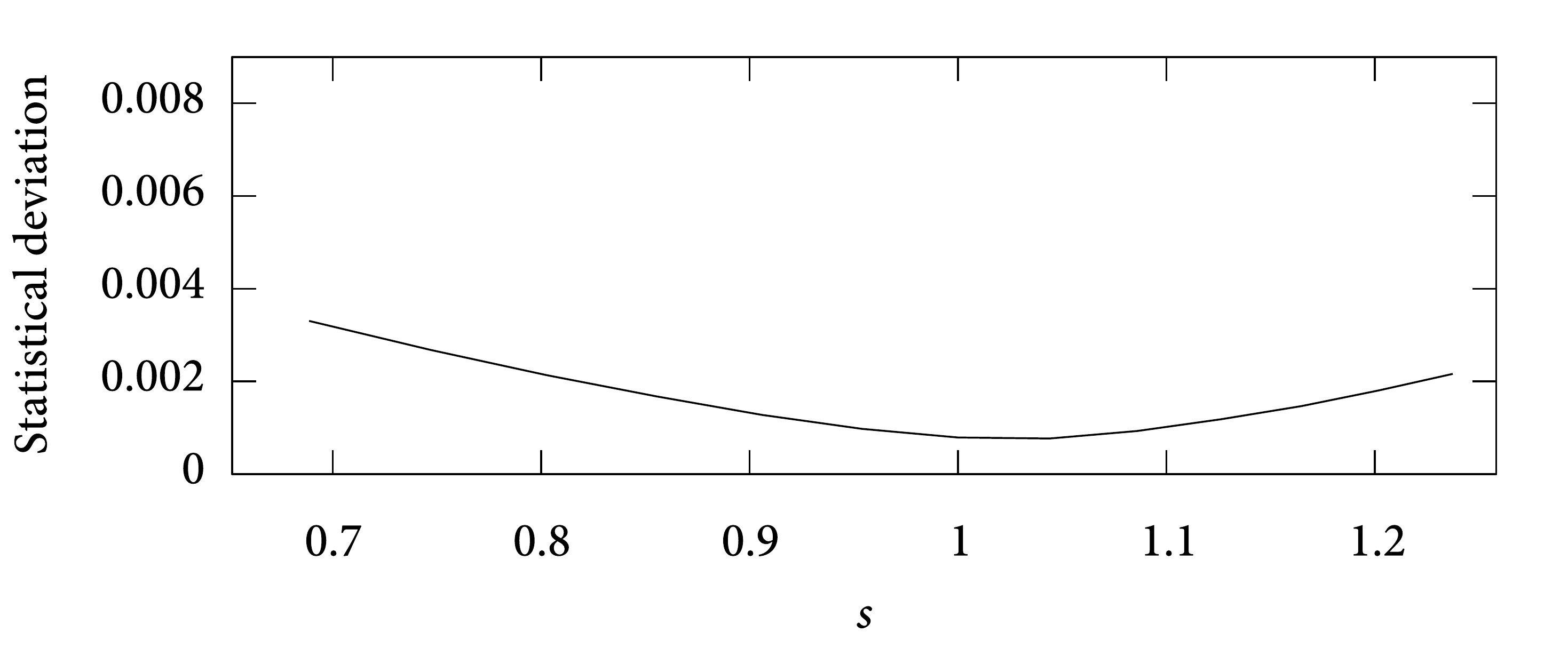}
    \caption{Average statistical deviation from the average energy distribution for 180~air showers of different energy and primary species, averaged in slant depth~(top), relative evolution stage~(middle), and age~(bottom). On average, the longitudinal range is the same in each plot.}
    \label{fig:depth_vs_age}
\end{figure}

As an example of this comparison we show in Fig.~\ref{fig:depth_vs_age} the statistical deviation from the mean energy distribution at each level. Plots are drawn as a function of~$t$ and their corresponding values in $X$ and~$s$. For descriptions in $t$ and~$s$, universality is highest near the shower maximum, because at that point all showers are at the same evolutionary stage by definition. This does not apply to the description in slant depth, where the shower maxima are not lined up. In this case, the relatively fast evolution for younger showers is reflected in falling deviations with depth. When the deviation is plotted for other physical quantities such as momentum angle or lateral distance, all curves behave in a similar manner as in Fig.~\ref{fig:depth_vs_age}.

Showers described in terms of~$X$ are less universal than those described in $s$ or~$t$, and slant depth is therefore rejected as parameter of choice. Between the two remaining descriptions, the difference is much smaller. Universality is slightly better for descriptions in evolution stage $t$ for $t>-8$, though the difference is insignificant. For very young showers $s$~is a better description, but this stage is not of interest observationally because the number of particles is so small. Comparing longitudinal shower size profiles, if showers are compared at the same evolution stage~$t$, better universality is found than when shower age~$s$ is used~\citep{2008:Muller}. Therefore, we describe electron and positron distributions in terms of relative evolution stage~$t$ in this work.

%
%
%
%
The total number of particles in the air shower crossing a plane at level~$t$ perpendicular to  the primary's trajectory is~$N(t)$. We define
\begin{equation}
    N(t;\mu)
        \equiv \pardiff{N(t)}{\mu}
    \quad\textrm{and}\quad
    n(t;\mu)
        \equiv \frac{1}{N(t)}\pardiff{N(t)}{\mu}
\end{equation}
as, respectively, the total and the normalised differential number of particles with respect to some variable~$\mu$. Likewise, distributions as a function of two variables $\mu$ and~$\nu$ are defined as
\begin{equation}\label{eq:dist_def}
    N(t;\mu,\nu)
        \equiv \pardiff{^2N(t)}{\mu\,\partial\nu}
    \quad\textrm{and}\quad
	n(t;\mu,\nu)
        \equiv \frac{1}{N(t;\mu)}\pardiff{^2N(t)}{\mu\,\partial\nu},
\end{equation}
with dimension~$[\mu\nu]^{-1}$ and $[\nu]^{-1}$, respectively. Note that the definition of~$n(t;\mu,\nu)$ implies that the distribution is normalised by integrating only over the last variable:
\begin{equation}\label{eq:dist_norm}
    \int_{\nu_\min}^{\nu_\max}n(t;\mu,\nu)\,\d\nu=1,
\end{equation}
making the normalisation independent of~$\mu$. In this expression, $\nu_\min$ and~$\nu_\max$ are the minimum and maximum values up to which the histograms are calculated.

The distributions~$n(t;\mu,\nu)$ presented in the following sections may be used to obtain realistic energy-dependent particle densities for an air shower, if the values of $X_\max$ and~$N_\max$ are given. One needs only to calculate the total number of particles~$N(t)$ at the desired shower evolution stage. An estimate of $N(t)$ can be obtained directly from shower profile measurements or through one of the many parameterizations available~\citep{1960:Greisen,1978:Gaisser,2001:Abu-Zayyad,2001:Pryke}.

%
%
\section{Energy spectrum}
%
%
\begin{figure}
    \includeplaatje[width=\figurewidth]{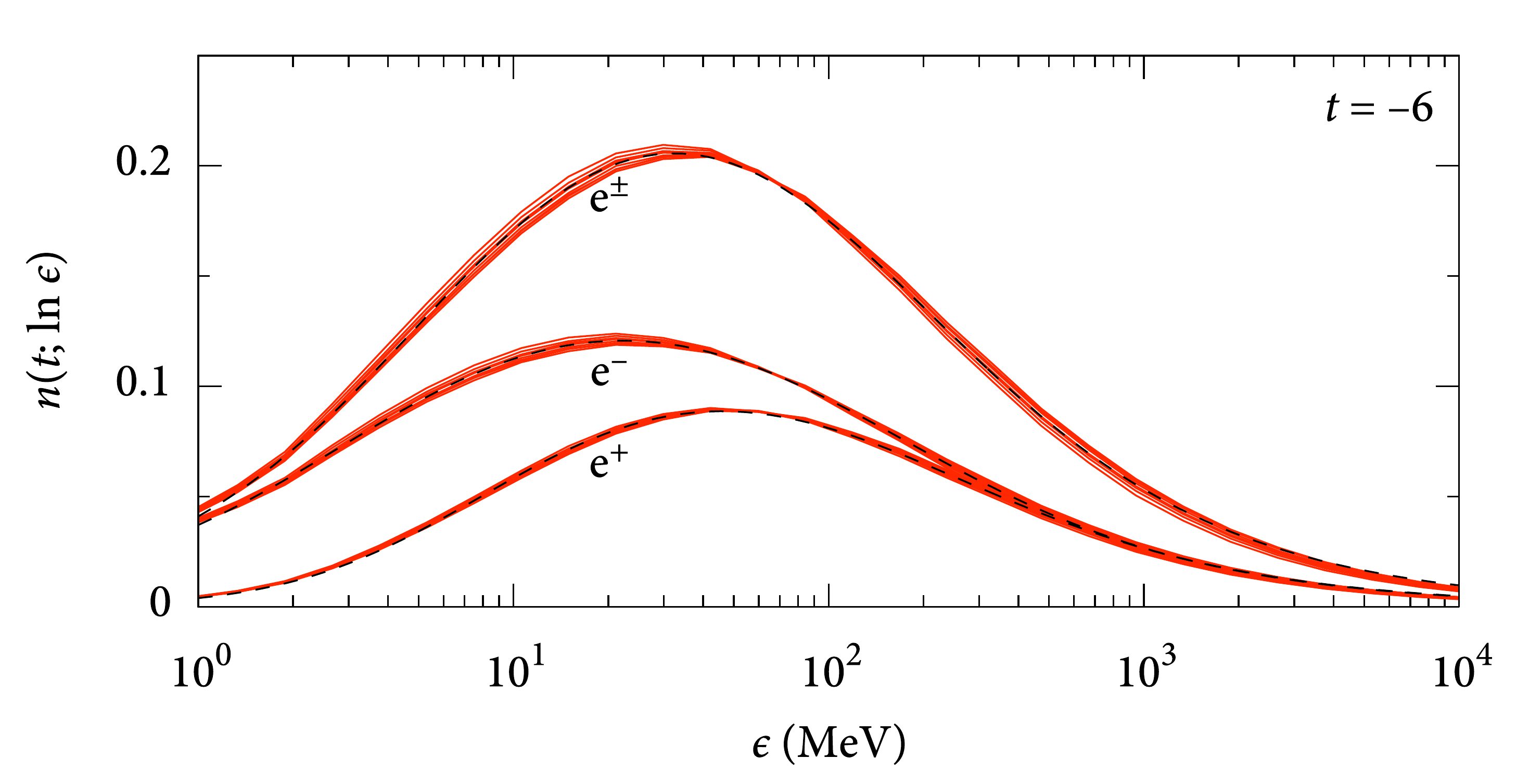}\\[-.11\figurewidth]
    \includeplaatje[width=\figurewidth]{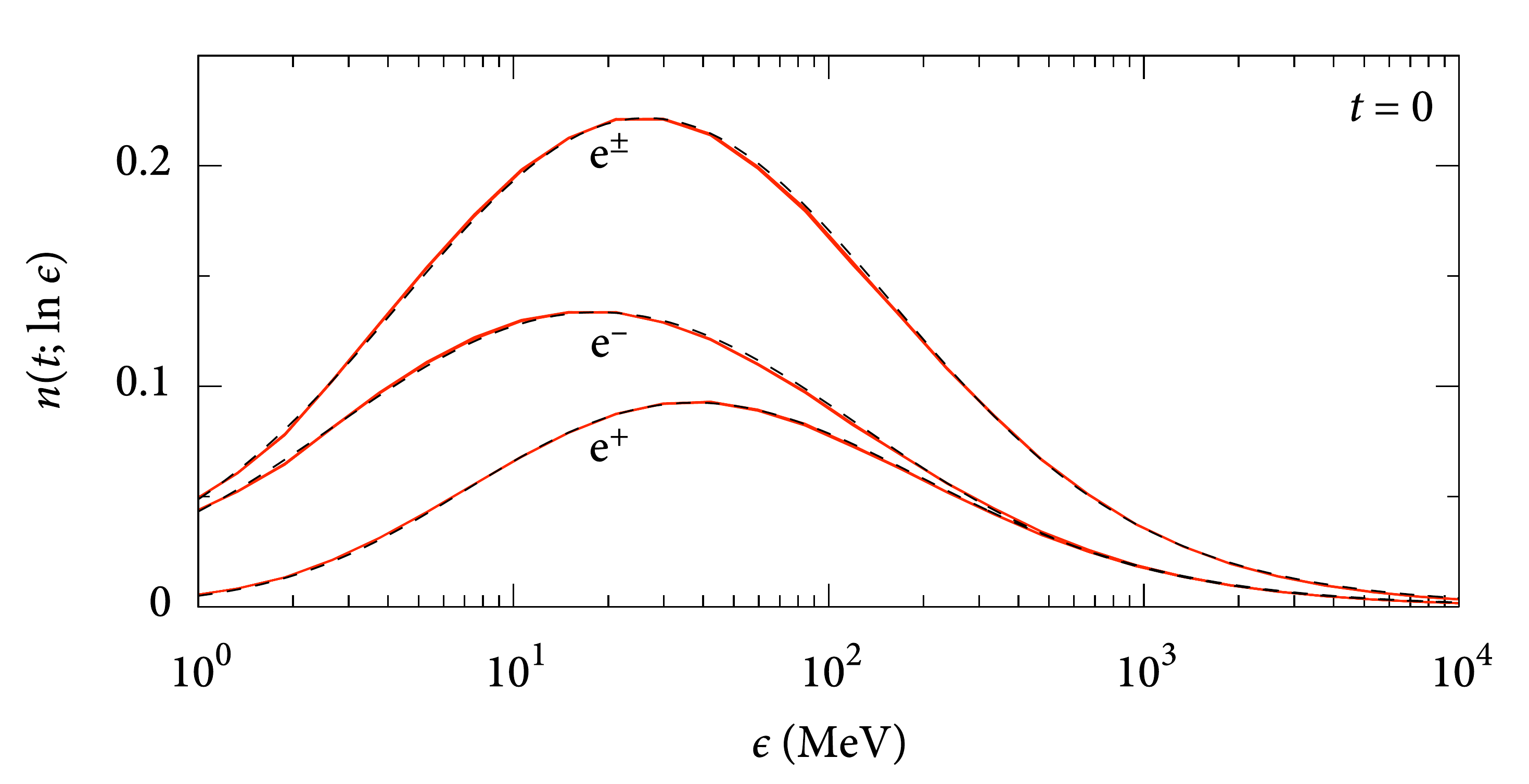}\\[-.11\figurewidth]
    \includeplaatje[width=\figurewidth]{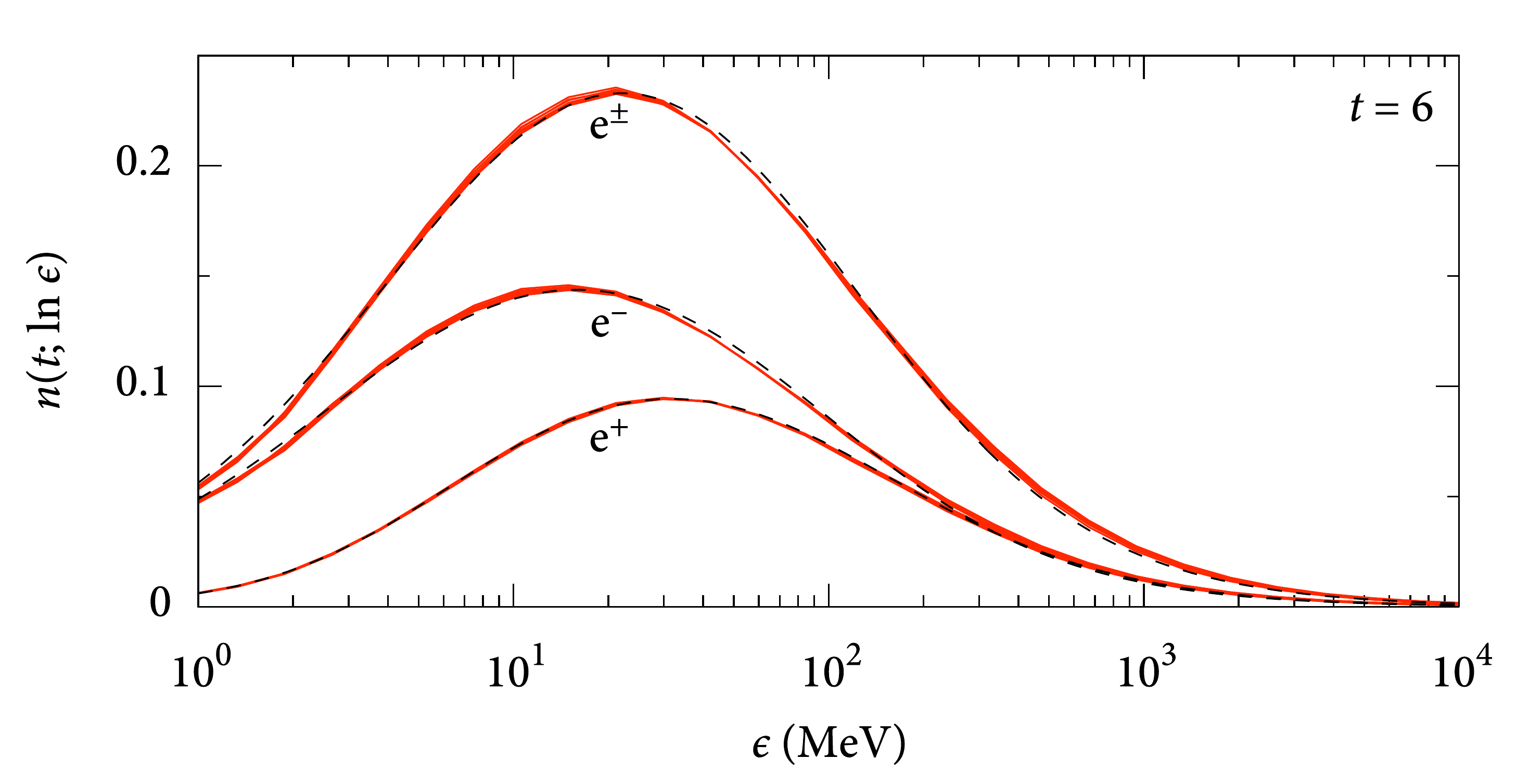}
    \caption{Average energy distribution for different evolution stages $t=-6,0,6$ for electrons (marked~$\lepton^-$), positrons~($\lepton^+$), and their sum~($\lepton^{±}$). Background curves represent simulated distributions for different primaries (p, Fe, and~$\photon$) and energies ($10^{17}$, $10^{18}$ and $10^{19}$\unit{eV}). The corresponding parameterized distributions from~\eqref{eq:energy_t} are plotted on top (dashed).}
    \label{fig:energy-fit}
\end{figure}
From cascade theory, the energy spectrum of electrons and positrons as a function of shower age takes an analytical form as derived by \citet{1941:Rossi}; a thorough previous study of this parameterization was done by \citet{2006:Nerling}. Loosely translating this description in terms of~$t$, we replace the equation by
\begin{equation}\label{eq:energy_t}
    n(t;\ln\epsilon)
         = \frac{A_0\epsilon^{\gamma_1}}
                {(\epsilon+\epsilon_1)^{\gamma_1}
                 (\epsilon+\epsilon_2)^{\gamma_2}},
\end{equation}
where~$\epsilon$ is the energy of a given secondary particle in the shower, and $\epsilon_{1,2}$ depend on~$t$. We have performed a fit to this function for electrons, positrons and their sum, indirectly providing a description of the negative charge excess of extensive air showers as a function of evolution stage and secondary energy. In these fits the exponent~$\gamma_1$ was fixed at $\gamma_1=2$ for positrons and $\gamma_{1}=1$ for both electrons and the total number of particles. The parameters for all three cases are explained in Appendix~\ref{app:energy}.

When applied to \corsika\ showers initiated by different species at different energies, the energy distribution~\eqref{eq:energy_t} is reconstructed accurately. This is shown in Fig.~\ref{fig:energy-fit}, where the simulated energy distributions are compared to their parameterizations for evolution stages $t=-6,0,6$. For shower stages $-6<t<9$, in the energy region $1\unit{MeV}<\epsilon<1$\unit{GeV}, which is most relevant for observation of geosynchrotron or Cherenkov radiation, deviations are generally smaller than~$10$\unit{\%} and never exceed~$25$\unit{\%} for all three parameterizations. For very young showers (Fig.~\ref{fig:energy-fit}, top panel), increasing deviations are mainly caused by variations in primary energy, not by primary species type. Therefore, it highlights a diminished accuracy to universally describe showers at $t<-6$ rather than hadronic model-dependence.

Using~\eqref{eq:energy_t}, a similar level of universality of the energy distribution of electrons and positrons is reached as previously obtained with a description in~$s$~\citep{2006:Nerling}. This basic observation is an important one, as it allows us to study other physical quantities in dependence of the electron energy in the remainder of this work.

%
%
\section{Angular spectrum}
       \label{sec:vertmom}
%
%
\begin{figure}
    \centerline{\includeplaatje[width=\figurewidth]{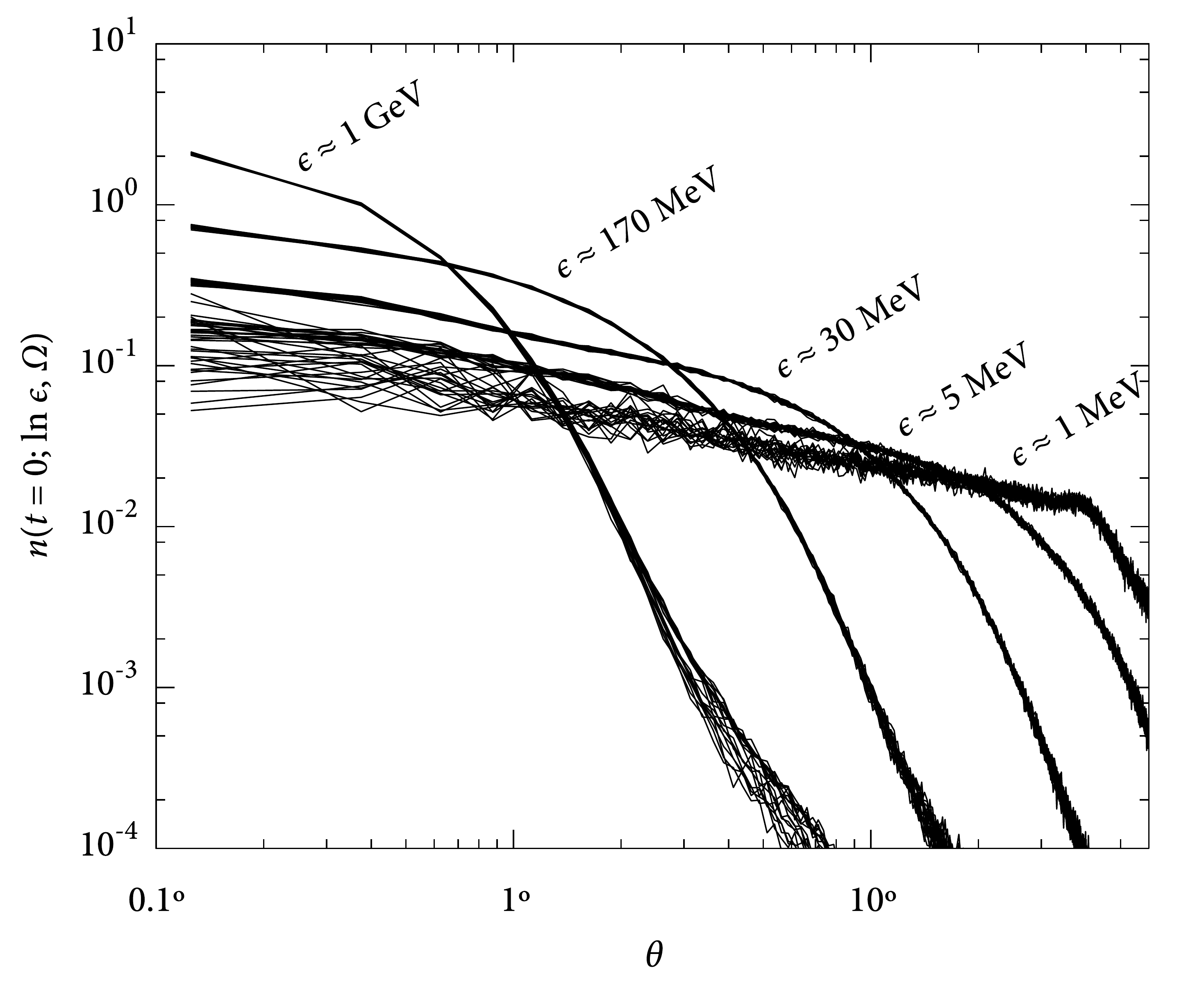}}
    \caption{Electron distributions~$n(t=0;\ln\epsilon,\Omega)$ at different electron energies as a function of momentum angle to the shower axis for 20 individual showers initiated by $10^{18}$\unit{eV} protons. $0$º~is along the primary's trajectory, $90$º~is perpendicular to the shower axis.}
    \label{fig:dist_theta_ind}
\end{figure}
The angular distribution of particles is an important factor for observations with Cherenkov and radio telescopes. For successful radio detection an antenna needs to be placed close to the shower impact position, because geosynchrotron radiation is beamed in a very narrow cone in the direction of propagation~\citep{2003:Huege:A&A}. As far as the particle distributions are concerned, the size of the patch that is illuminated on the ground then depends on the lateral distribution of the particles (cf. Sect.~\ref{sec:lateral}) and the angle with respect to the shower axis at which they propagate. Likewise, for Cherenkov observations the angle at which photons are emitted is a convolution of the density-dependent Cherenkov angle, which is of the order of~$\sim1$º, and the angular distribution of the particles that emit them.

Fig.~\ref{fig:dist_theta_ind} shows the angular distribution of particles as simulated in 20~individual vertical proton showers at $10^{18}$\unit{eV} as a function of~$\theta$. To compensate for the increase in solid angle with rising~$\theta$, the distribution of vertical momentum angles plotted here is defined in terms of~$\Omega$ as
\begin{equation}
n(t;\ln\epsilon,\Omega)=\frac{n(t;\ln\epsilon,\theta)}{\sin\theta}.
\end{equation}
Since the majority of all electrons and positrons stays close to the shower axis, we focus on this part of the distribution. We will ignore the more horizontal part further away from the axis that can be seen at the right end of the curve for~$1$\unit{GeV} in Fig.~\ref{fig:dist_theta_ind}. When~$\theta$ is plotted on a logarithmic scale, it becomes clear that there is a plateau close to the shower axis at all energies and a sharp drop at a certain angle that depends on secondary energy.

\begin{figure}
    \includeplaatje[width=\figurewidth]{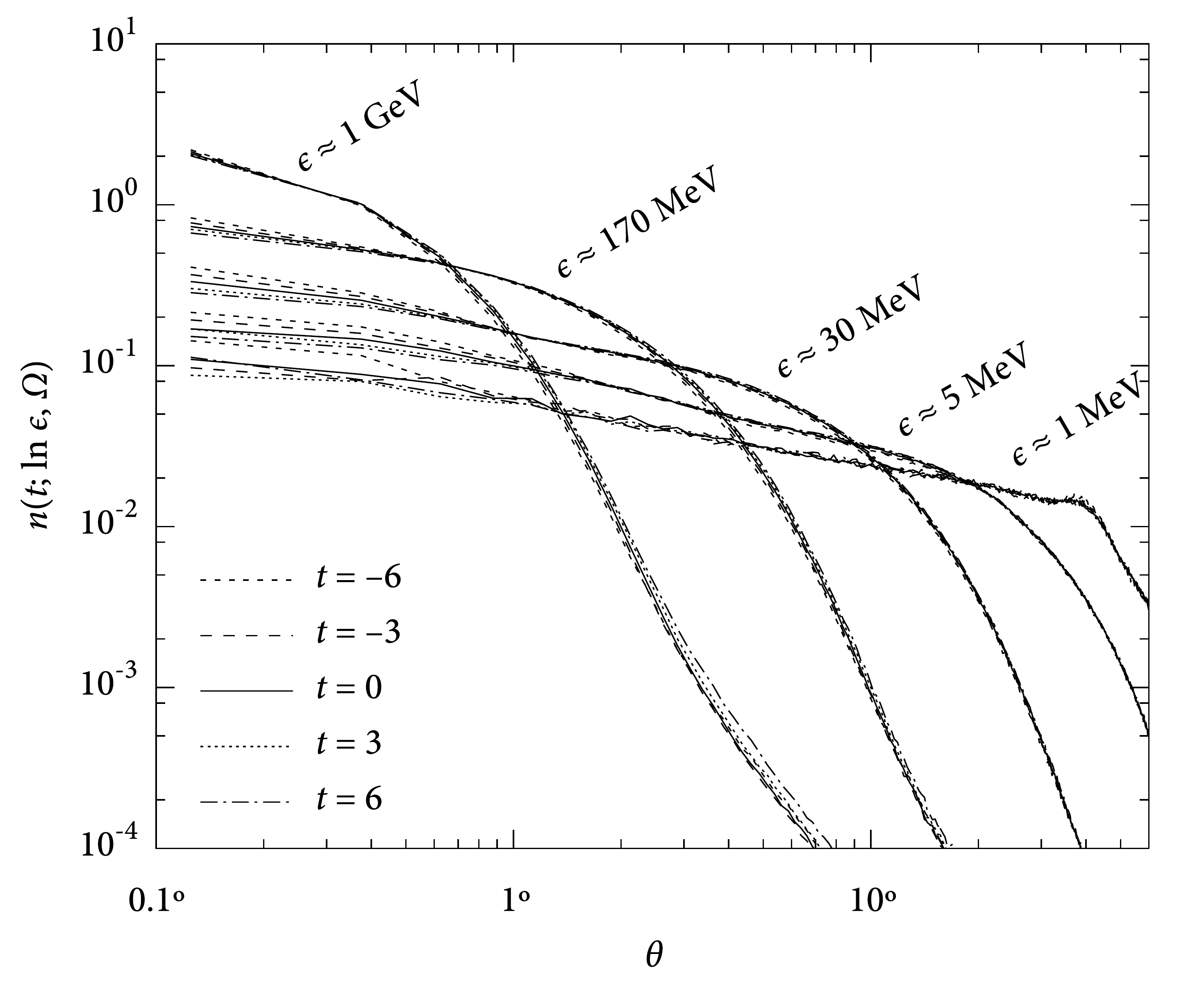}
    \caption{Normalised average distributions~$n(t;\ln\epsilon,\Omega)$ for different shower stages, averaged over 20~proton-initiated showers at $10^{18}$ eV.}
    \label{fig:dist_theta_dep}
\end{figure}
Fig.~\ref{fig:dist_theta_dep} extends the angular distributions to different shower stages. The differences in the distributions are clearly smaller than the differences between individual showers, as noted earlier~\citep{2006:Nerling,2005:Giller,2005:Giller:proc}. The differential electron distribution with regard to the direction of the particle's momentum is therefore independent of shower stage. In addition, no perceptible dependence on incidence zenith angle or primary energy was found. When looking at different primary species, universality seems somewhat less convincing: spectra for heavier primary species tend to be wider at higher electron energies. The effect is too small, however, to be of consequence in our analysis.

\begin{figure}
    \includeplaatje[width=\figurewidth]{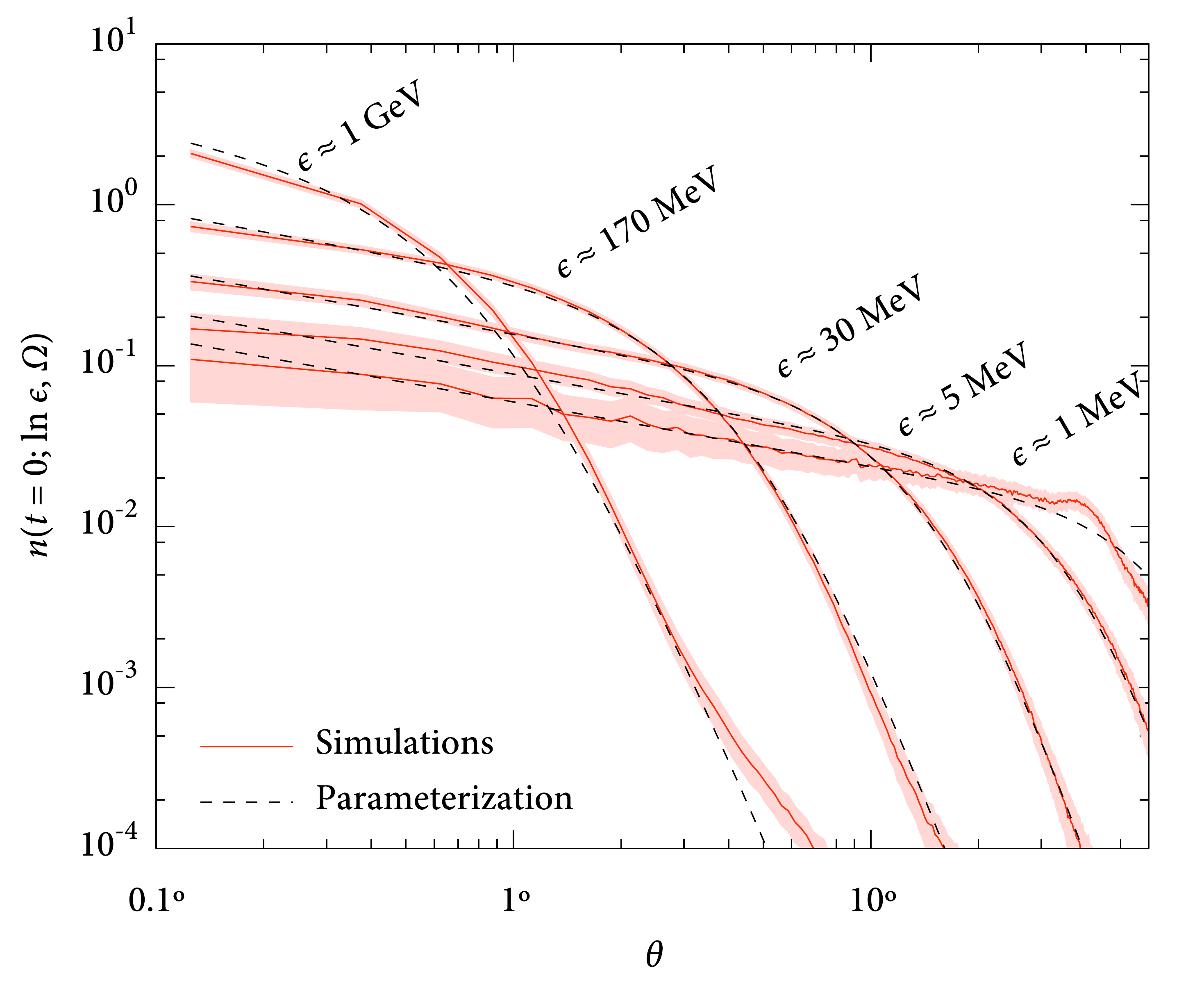}
    \caption{Normalised average electron distributions~$n(t=0;\ln\epsilon,\Omega)$ (solid) for 20~proton showers at $10^{18}$~eV with $3\sigma$~statistical error margins (filled area). For each energy, corresponding parameterizations according to~\eqref{eq:vertmom} are also drawn (dashed).}
    \label{fig:dist_theta_fit}
\end{figure}
The universality with respect to~$t$ allows us to parameterize this distribution as a function of two physical quantities only: momentum angle and energy. We propose the form
\begin{equation}\label{eq:vertmom}
    n(t;\ln\epsilon,\Omega)
         = C_0
           \left[\left(\E{b_1}\theta^{\alpha_1}\right)^{-1/\sigma}
             +   \left(\E{b_2}\theta^{\alpha_2}\right)^{-1/\sigma}
           \right]^{-\sigma},
\end{equation}
to describe the distribution. Values for $\alpha_i$ and~$b_i$, which envelop the dependence on~$\epsilon$, are chosen such that the first term describes the flatter portion of the angular distribution parallel to the shower axis and the second represents the steep drop. The value of~$\sigma$ determines the smoothness of the transition from the flat region to the steep region. Best fit values for $\sigma$, $b_i$, and~$\alpha_i$ are given in Appendix~\ref{app:vertical}. The dependence of these parameters on the secondary energy~$\epsilon$ was determined purely empirically. For several energies, the parameterized forms are plotted along with their associated simulated distributions in Fig.~\ref{fig:dist_theta_fit}, showing good correspondence between the two. The parameterization provides a good description of the simulated distribution for the energy region $1\unit{MeV}<\epsilon<10$\unit{GeV} and $\theta<60$º.

\begin{figure}
    \includeplaatje[width=\figurewidth]{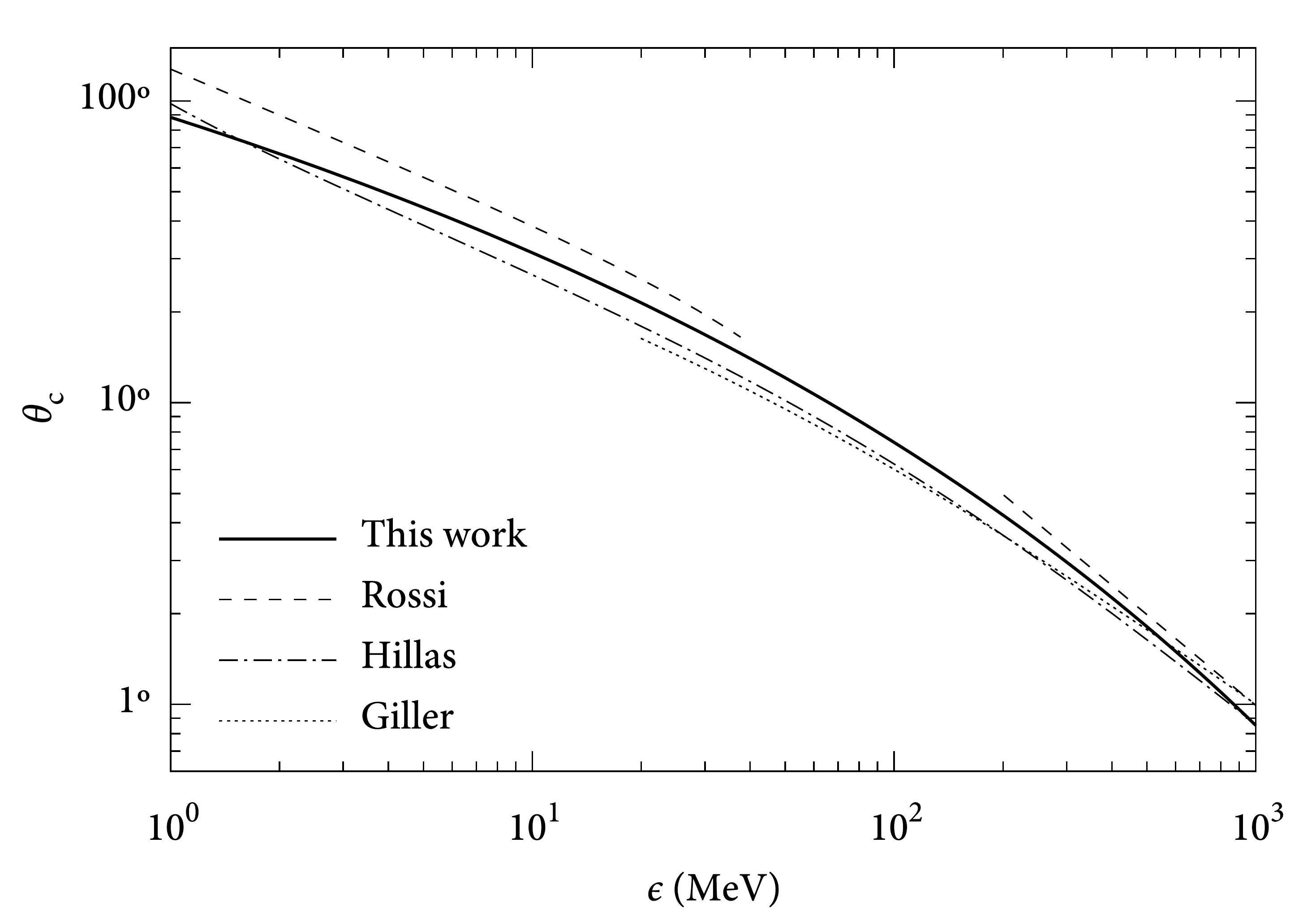}
    \caption{Cutoff angle~$\theta_\mathrm{c}$ according to~\eqref{eq:theta_cutoff} for the angular distribution as a function of secondary energy (solid line). Also shown are theoretical predictions for~$\theta_\textsc{rms}$ from~\citet{1941:Rossi} (dashed) as well as empirical relations from~\citet{1982:Hillas:1} (dash-dotted) and~\citet{2005:Giller} (dotted).}
    \label{fig:theta_cutoff}
\end{figure}
We now define the cutoff angle~$\theta_\mathrm{c}$ as one half of the angle at which $\E{b_1}\theta^{\alpha_1}=\E{b_2}\theta^{\alpha_2}$:
\begin{equation}\label{eq:theta_cutoff}
    \theta_\mathrm{c}(\epsilon) = \frac{1}{2}\exp\left[
            -\frac{b_1-b_2}
                  {\alpha_1-\alpha_2}
        \right].
\end{equation}
For high energies, where the momentum angle is smaller than~$90$º for the majority of particles, $\theta_\textrm{c}$~is a measure for the root mean square value~$\theta_\textsc{rms}$ of the particle momentum angles. This is outlined in Fig.~\ref{fig:theta_cutoff}, in which~$\theta_\mathrm{c}$ is plotted as a function of energy. Theoretical root mean square scattering angles according to~\citet{1941:Rossi} in high and low secondary energy limits are also drawn, as well as empirical models as parameterized in~\citet{1982:Hillas:1} and~\citet{2005:Giller}. At high energies, the theoretical average scattering angle is expected~$\propto\epsilon^{-1}$, while at low energies it is~$\propto\epsilon^{-1/2}$. This behaviour is reproduced properly for the cutoff angle. For low secondary energies ($\epsilon≲3$\unit{MeV}), the definition of a cutoff or root mean square angle becomes inapplicable as the angular distribution widens, covering all angles. For $\epsilon>2$\unit{MeV}, no appreciable difference was found between the angular distributions of positrons on the one hand and electrons on the other.

Because our histograms do not have any sensitivity in the azimuthal direction by design, no dependence on the geomagnetic field could be determined. Previous work has shown that the effect on the angular distribution is probably small, but not negligible~\cite{1982:Hillas:1,1983:Elbert}. Because the accuracy of simulations has rather improved since these studies were carried out, it would be worthwhile to investigate the effect of the geomagnetic field in greater detail.
%
%
\section{Outward momentum distribution}
%
%
Let us define~$\phi$ as the angle of a particle momentum vector projected in the plane perpendicular to the shower axis with respect to the outward direction, such that $\phi=0$º for a particle moving away from the shower axis, and $\phi=180$º for a particle moving towards it. We will refer to this angle as the horizontal momentum angle. The effect of fluctuations in the horizontal angular distribution is generally much less important than those in the vertical angular spectrum. In fact, the distribution of the $\phi$~angle of the particles does not have any influence on the observed signal when the distance from the observer to the shower is much larger than the average distance from the shower particles to the shower axis, as is the case in air fluorescence observations. This is because the cylindrical symmetries of the momentum angles and the shower geometry cancel out independently of the shape of the distribution. Geosynchrotron radiation, however, will only produce a significant signal reasonably close to the shower axis, because the shower front is thicker in length further away (cf. Sect.~\ref{sec:front}), breaking down coherence. Therefore, the horizontal momentum angle spectrum has to be taken into account for radio measurements.

\begin{figure}
    \centerline{\includeplaatje[width=\figurewidth]{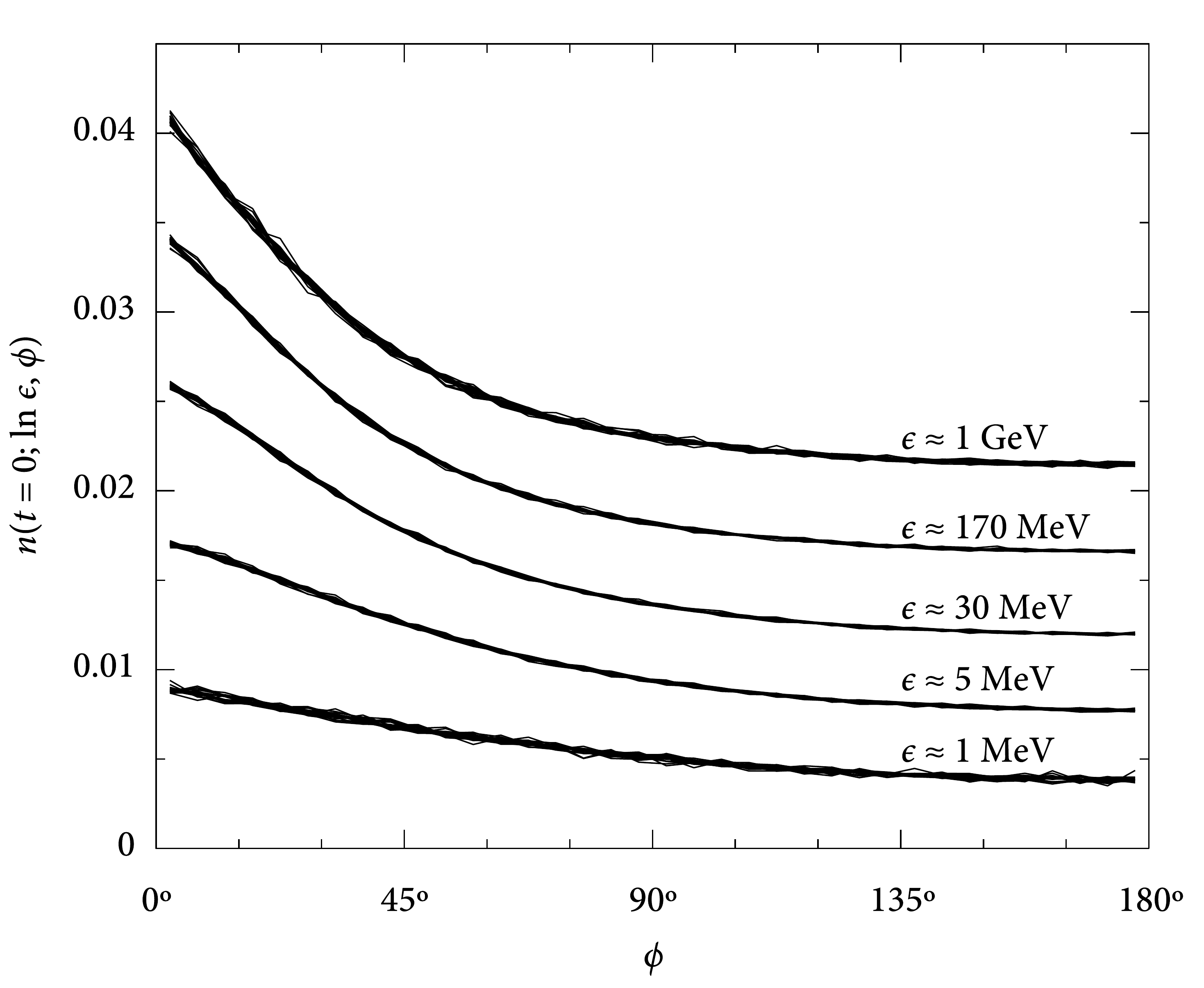}}%
    \caption{Normalised simulated horizontal angular electron distributions for 20~individual showers initiated by $10^{18}$~eV protons at different energies. Consecutive curve sets are shifted up by~0.005 to distinguish them better; curves for $1$\unit{MeV} are at the actual level.}
    \label{fig:dist_phi_ind}
\end{figure}
Simulated distributions~$n(t;\ln\epsilon,\phi)$ at $t=0$ are plotted in Fig.~\ref{fig:dist_phi_ind} for the reference set. We observe that high-energy particles tend to move outward more than lower-energy particles. This can be explained by considering the collisions in which high-energy electrons and positrons are created, as they primarily occur close to the shower axis. Hence reaction products are transported away from the shower core due to their transverse momenta. Electrons and positrons with lower energies, on the other hand, are also created further away from the shower core.

 No significant dependencies on incident zenith angle, primary energy, and primary species were found, so the horizontal momentum angular spectra are universal. Additionally, the shape of the distribution does not change significantly for $\epsilon>2$\unit{MeV} when only electrons or only positrons are considered. There is some dependence in terms of~$t$, however: the distribution appears to soften with evolution stage.
This effect can be explained from the expanding spatial structure of the shower with age.

\begin{figure}
    \includeplaatje[width=\figurewidth]{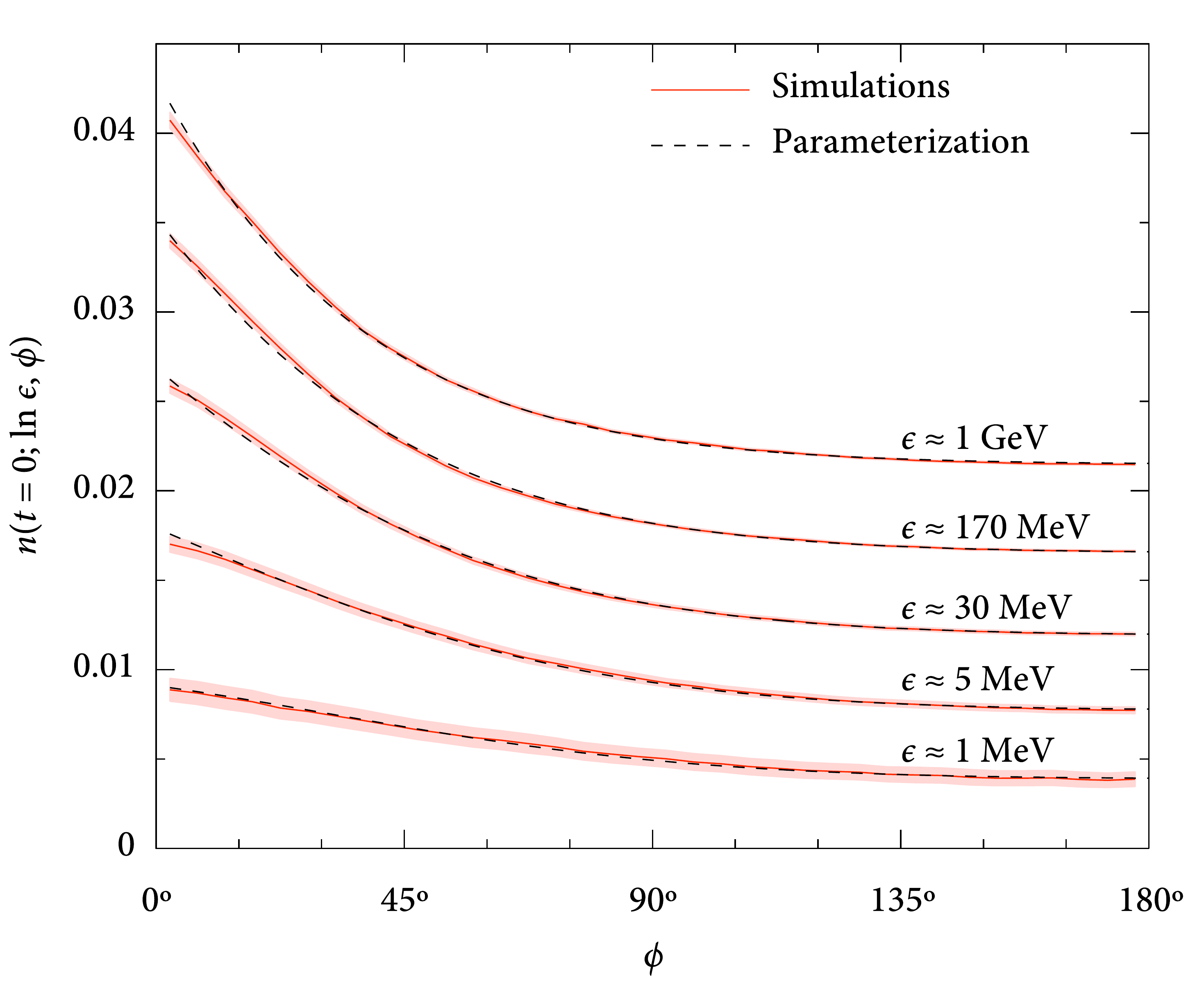}
    \caption{Normalised average electron distributions~$n(t=0;\ln\epsilon,\phi)$ (solid) for 20~proton showers at $10^{18}$\unit{eV} with $3\sigma$~statistical error margins (filled area). For each energy, corresponding parameterizations according to~\eqref{eq:hormom} are also drawn (dashed).}
    \label{fig:dist_phi_fit}
\end{figure}
The distribution of $n(t;\phi)$ is very nearly exponential for electrons and positrons with energies over $10$\unit{GeV}, while it has a slight bulge around the outward direction at lower energies. To describe the distribution, we use the parameterization
\begin{equation}\label{eq:hormom}
    n(t;\ln\epsilon,\phi)
         = C_1[1+\exp(\lambda_0-\lambda_1\phi-\lambda_2\phi^2)],
\end{equation}
a form which accurately reproduces the distribution. The resulting parameter values $\lambda_0(t,\epsilon)$, $\lambda_1(\epsilon)$, and~$\lambda_2(\epsilon)$ are explained in Appendix~\ref{app:horizontal}. The reference set, drawn together with its corresponding parameterization in Fig.~\ref{fig:dist_phi_fit}, shows a high level of agreement. For other shower parameters and stages, there is a similar degree of consistency.

%
%
\section{Lateral distribution}
       \label{sec:lateral}
%
%
The lateral spread of particles in an air shower is of direct relevance since it is the primary means of obtaining information about the shower in ground-based scintillator experiments measuring particle densities at different lateral distances. By integrating the measured distribution or using the particle density at a given distance, an estimate for the primary energy can be made. Exact knowledge of the lateral distribution shape is therefore crucial to accurately determine the shape of the cosmic-ray energy spectrum.

\begin{figure}
    \centerline{\includeplaatje[width=\figurewidth]{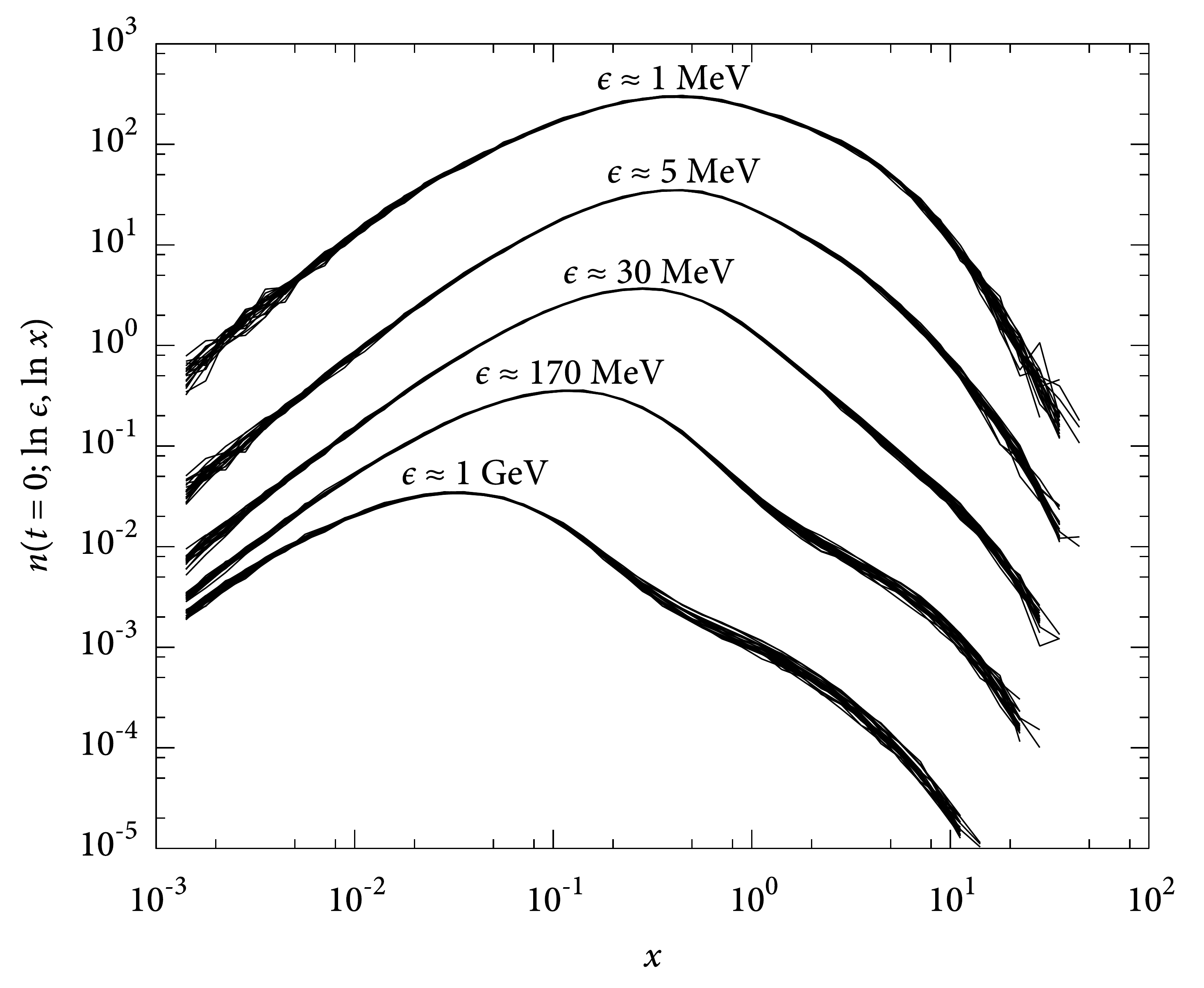}}%
    \caption{Electron distributions~$n(t=0;\ln\epsilon,\ln x)$ for different electron energies as a function of distance to the shower axis for 20 individual showers initiated by $10^{18}$\unit{eV} protons. The curve set for $1$\unit{GeV} is at the actual level; consecutive sets are shifted up by a factor of~10.}
    \label{fig:dist_r_ind}
\end{figure}
When looking at the lateral distribution of electron and positrons in terms of the lateral distance~$r$ from the shower axis, a very poor level of universality is encountered. This is mainly due to differences in atmospheric density at the individual values of~$X_\max$. We can compensate for these differences by expressing the lateral distance in terms of the Molière unit~$r_\mathrm{M}$, defining~\citep{2003:Dova}
\begin{equation}
    x\equiv\frac{r}{r_\mathrm{M}}\simeq\frac{r\rho_\mathrm{A}(h)}{9.6\unit\gcm},
\end{equation}
where~$\rho_\mathrm{A}(h)$ is the atmospheric density as a function of height~$h$. For different values of~$\epsilon$, the normalised lateral particle distribution at $t=0$ is shown in Fig.~\ref{fig:dist_r_ind} as a function of distance for 20~individual proton showers. In this figure, all curves line up as the compensation for density is applied. Note that the physical density $N(t;r)$, expressed in particles per unit area, is proportional to $N(t;\ln x)/x^2$:
\begin{equation}
    N(t;\ln x)=\pardiff{N(t)}{\ln x}=2\pi x^2r_\mathrm{M}^2\frac{\d N(t)}{2\pi r\,\d r},
\end{equation}
and decreases strictly with distance from the shower axis.
As expected, particles with higher energies tend to remain closer to the shower axis. This agrees with the observation that the angle of their momentum to the shower axis is smaller.

\begin{figure}
    \centerline{\includeplaatje[width=\figurewidth]{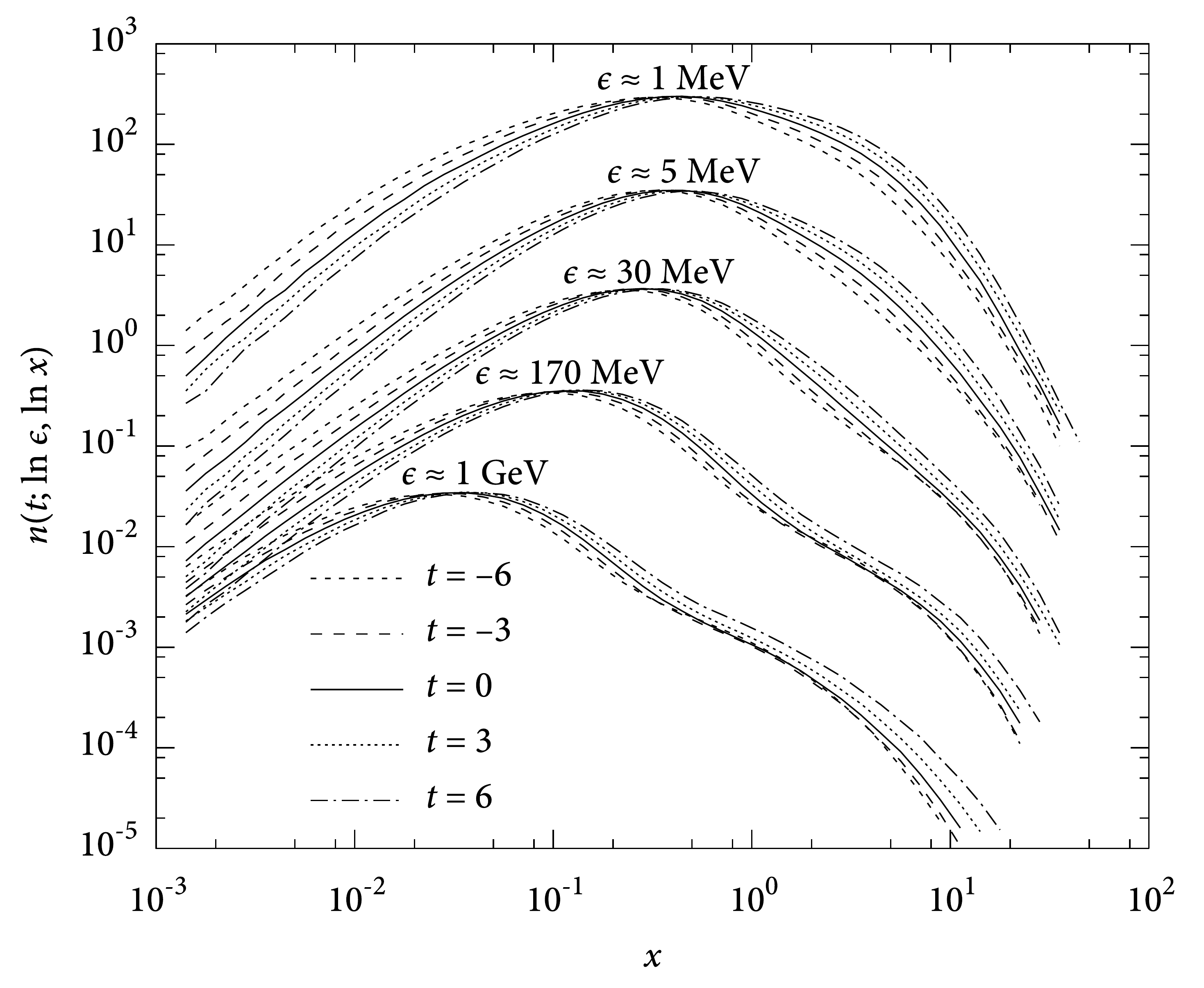}}%
    \caption{Average distributions~$n(t;\ln\epsilon,\ln x)$ for different shower stages, averaged over 20~proton-initiated showers at $10^{18}$\unit{eV}, clearly showing dependence on~$t$. Again, consecutive sets are shifted up by a factor of~10.}
    \label{fig:dist_r_dep_age}
\end{figure}
There is no statistically relevant dependence of the lateral distribution on zenith angle of incidence, nor does it change when electrons or positrons are considered separately, except at energies $\epsilon<10$\unit{MeV}. There is, however, a significant effect with shower stage as shown in Fig.~\ref{fig:dist_r_dep_age}: older showers tend to be wider at the same secondary energy. Therefore, unlike in the case of angular distributions, in any parameterization of the lateral distribution a dependence on~$t$ must be incorporated. There is also a minor effect of the energy of the primary on the distribution, but this is only appreciable for secondary energies of $\epsilon>1$\unit{GeV}.

\begin{figure}
    \includeplaatje[width=\figurewidth]{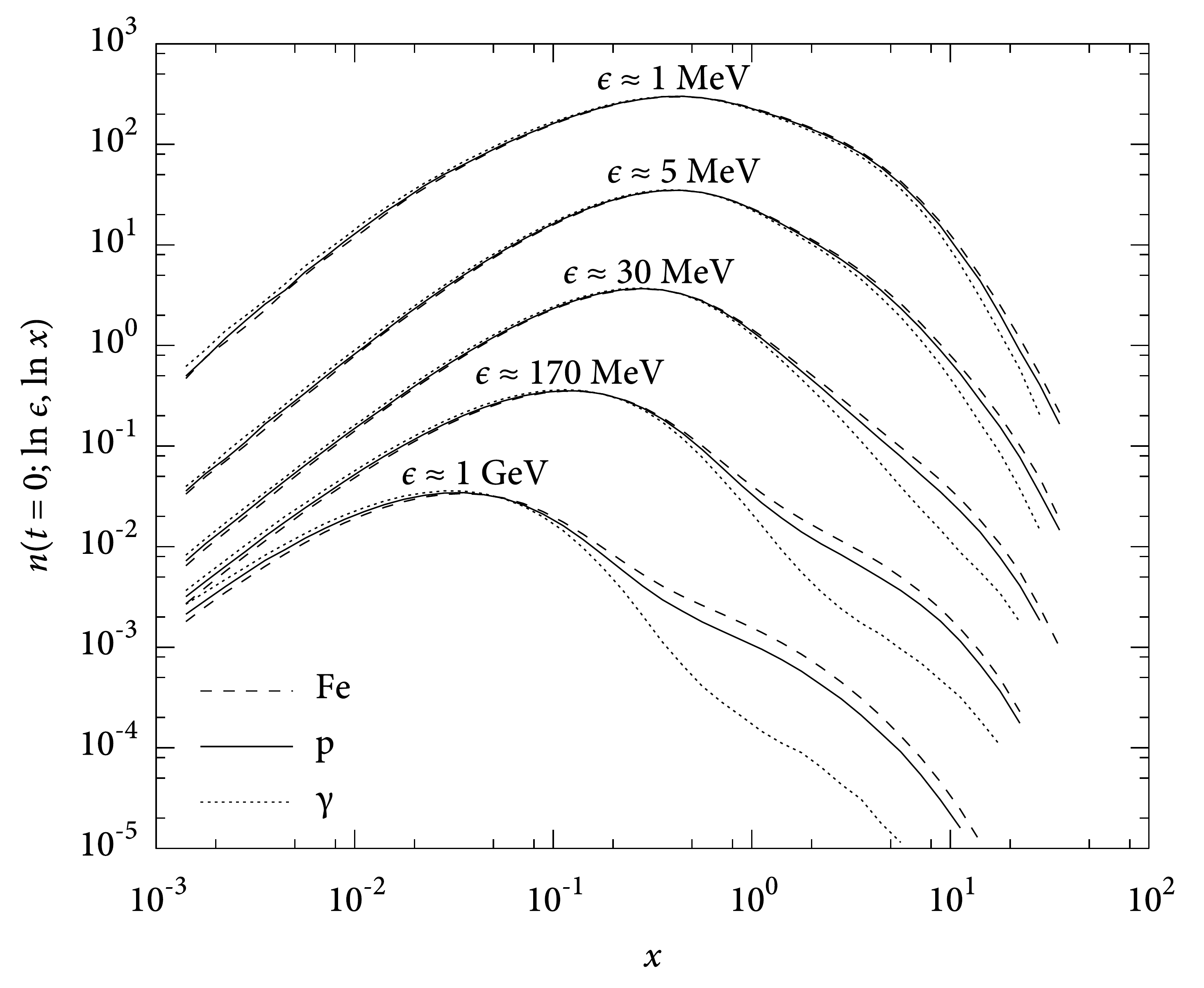}
    \caption{Average distributions~$n(t=0;\ln\epsilon,\ln x)$ for different primaries, averaged over 20~showers at $10^{18}$~eV. Again, consecutive sets are shifted up by a factor of~10. Note the dependence on species of the bulge on the right.}
    \label{fig:dist_r_dep_species}
\end{figure}
\begin{figure}
    \includeplaatje[width=\figurewidth]{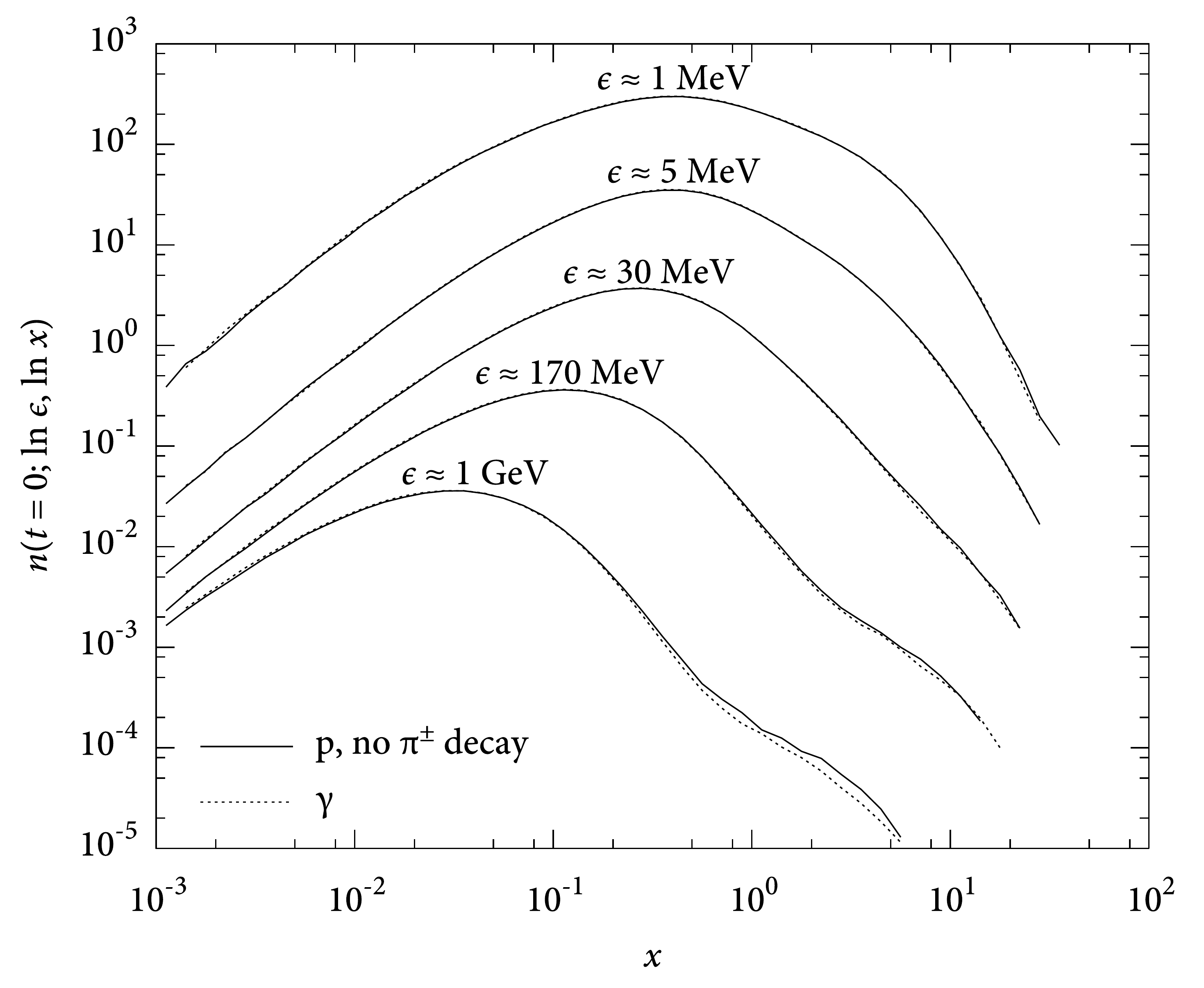}
    \caption{Comparison of average distributions~$n(t=0;\ln\epsilon,\ln x)$ at $10^{17}$\unit{eV} for 20~standard photon showers to 20~proton showers in which $\pion^{±}$ decay was disabled. Again, consecutive sets are shifted up by a factor of~10.}
    \label{fig:dist_r_pion_decay}
\end{figure}
From Figs.~\ref{fig:dist_r_ind}, \ref{fig:dist_r_dep_age}, and~\ref{fig:dist_r_dep_species} it is observed that each curve is a combination of two separate contributions. The left peak, the shape of which does not depend significantly on primary energy or species, is produced through the main electromagnetic formation channel of cascading steps of bremsstrahlung and pair creation. The second bulge shows a high level of dependence on primary species, as shown in Fig.~\ref{fig:dist_r_dep_species}. It tends to be less prominent for photon primaries, as for these species there is no significant contribution from the pion production channel. For hadronic primaries it is more significant, especially at higher secondary energies of $\epsilon>100$\unit{MeV}. The magnitude of the variation between different species does not change with~$t$, but its lateral position does slightly. The variations in strength of the second bulge for different primaries can be traced back to the contribution initiated by the decay channel $\pion^{±}→\muon^{±}+\neutrino_\muon$. This is shown in Fig.~\ref{fig:dist_r_pion_decay}, comparing a set of unaltered $10^{17}$\unit{eV} photon-initiated showers, which have no significant pion content, to a set of proton showers at the same energy in which the $\pion^{±}$~creation channel was disabled. Differences between their lateral distributions are smaller than statistical deviations.

This observation raises the question whether one could use this difference in lateral distribution to differentiate between primaries on an individual shower basis by their lateral distribution, independently of measurements of primary energy or depth of shower maximum. This would be a difficult task. First of all, appreciable difference in density only occurs at high energies and at some distance, implying that the total electron density in the region of sensitivity would be very small. Additionally, the effect does not appear at the same distance for different electron energies. This makes the feature less pronounced when an integrated energy spectrum is measured.

\begin{figure}
    \includeplaatje[width=\figurewidth]{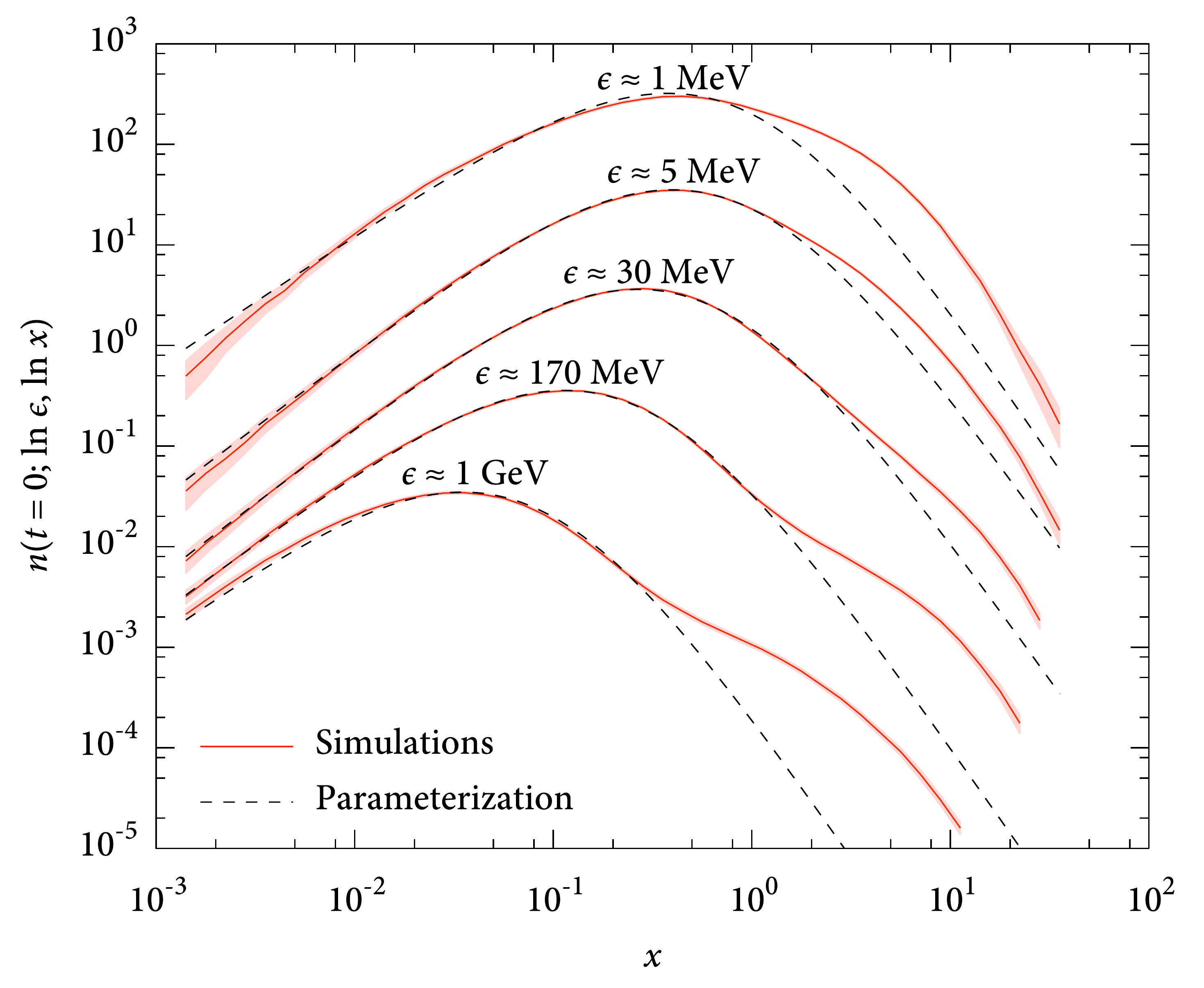}
    \caption{Normalised average electron distributions~$n(t=0;\ln\epsilon,\ln x)$ (solid) for 20~proton showers at $10^{18}$\unit{eV} with $3\sigma$~statistical error margins (filled area). For each energy, corresponding parameterizations according to~\eqref{eq:lateral3d} are also drawn (dashed). Consecutive sets are again shifted up by a factor of~10.}
    \label{fig:dist_r_fit}
\end{figure}
Traditionally, the integral lateral electron distribution is described by a an approximation of the analytical calculation of the lateral distribution in electromagnetic cascades, the Nishimura-Kamata-Greisen (\textsc{nkg}) function~\citep{1958:Kamata,1965:Greisen}. The integral lateral distribution for our simulated set of showers~$n(t;\ln x)\propto x^2\rho_\textsc{nkg}$ is reproduced well by a parameterization of this form, provided that we allow the parameters to be varied somewhat. Let us define
\begin{equation}\label{eq:lateral2d}
    n(t;\ln x)
        = C_2 x^{\zeta_0} (x_1+x)^{\zeta_1}.
\end{equation}
as parameterization. In the original definition, described in terms of shower age~$s$, we have $\zeta_0=s$, $\zeta_1=s-4.5$, and $x_1=1$. Our simulated lateral spectra closely follow the values $\zeta_0=0.0238t + 1.069$, $\zeta_1=0.0238t - 2.918$, and $x_1=0.430$ to an excellent level for $10^{-3}<x<10$.

To reproduce the main bulge in the energy-dependent lateral electron distributions, we propose a slightly different function. The second bulge will be ignored here since it is much lower than the primary bulge, and its relative height depends heavily on primary species as mentioned earlier. The proposed parameterization is the same as~\eqref{eq:lateral2d}:
\begin{equation}\label{eq:lateral3d}
    n(t;\ln\epsilon,\ln x)=C_2'x^{\zeta_0'}(x_1'+x)^{\zeta_1'},
\end{equation}
mimicking the behaviour of the \textsc{nkg} function, but now also varying the parameters with~$\epsilon$. Appendix~\ref{app:lateral} explains the values of $x_i'$ and~$\zeta_i'$. As an example of the fit, Fig.~\ref{fig:dist_r_fit} compares the parameterization to the average distribution for proton showers at their maximum. The proposed parameters adequately reproduce the main bulge of the lateral distribution in the energy range of $1\unit{MeV}<\epsilon<1$\unit{GeV} for distances~$x>2\e{-3}$ and evolution stages~$-6<t<9$.

Neglecting the second bulge results in a slightly overestimated overall value for the normalisation. The disregarded tail only constitutes a minor fraction of the total number of particles, however, especially at high energies. This fact becomes even more evident if one considers that the actual distribution is obtained by dividing by~$x^2$.

\begin{figure}
    \includeplaatje[width=\figurewidth]{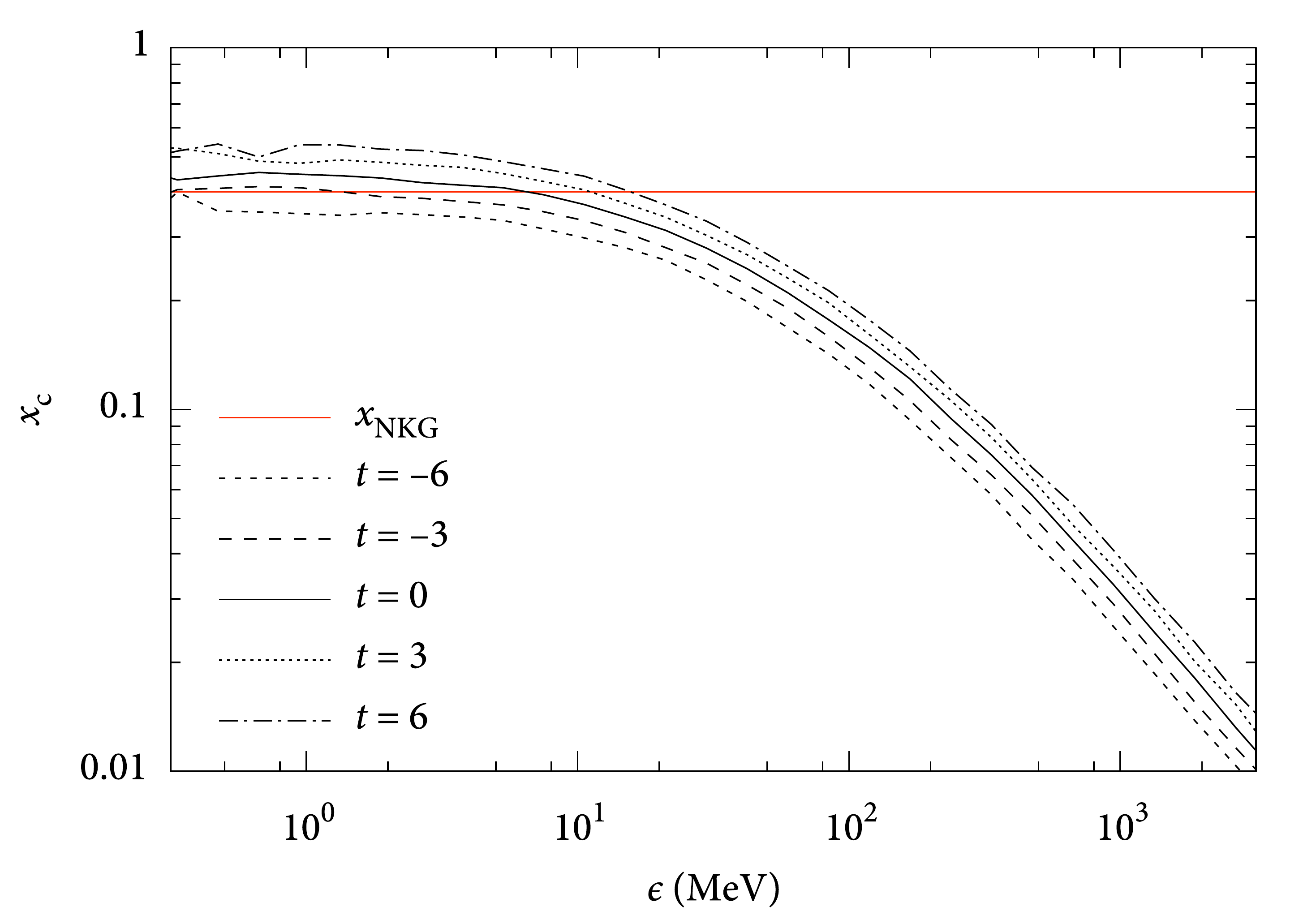}
    \caption{Cutoff distance~$x_\mathrm{c}$ as a function of secondary energy at different shower stages. The energy-independent overall break distance obtained from the \textsc{nkg} function is also plotted (horizontal line).}
    \label{fig:r_cutoff}
\end{figure}
The position of the break~$x_\mathrm{c}$, the distance of the highest peak in the distribution, is plotted in Fig.~\ref{fig:r_cutoff} for various shower stages for 20~averaged showers. The theoretical break distance from the original Nishimura-Kamata-Greisen distribution at the shower maximum, which is an integral distribution over all electron energies, is also plotted as a horizontal line. At lower energies, the two are in good agreement as expected.

%
%
\section{Delay time distribution}
       \label{sec:delay}
%
%
For radio geosynchrotron measurements the arrival time of charged particles is a vital quantity, because it determines the thickness of the layer of particles that form the air shower. This thickness in turn defines the maximum frequency up to which the resulting radio signal is coherent~\citep{2003:Huege:A&A,2008:Scholten}, which influences the strength of the radio signal on the ground.

\begin{figure}
    \centerline{\includeplaatje[width=\figurewidth]{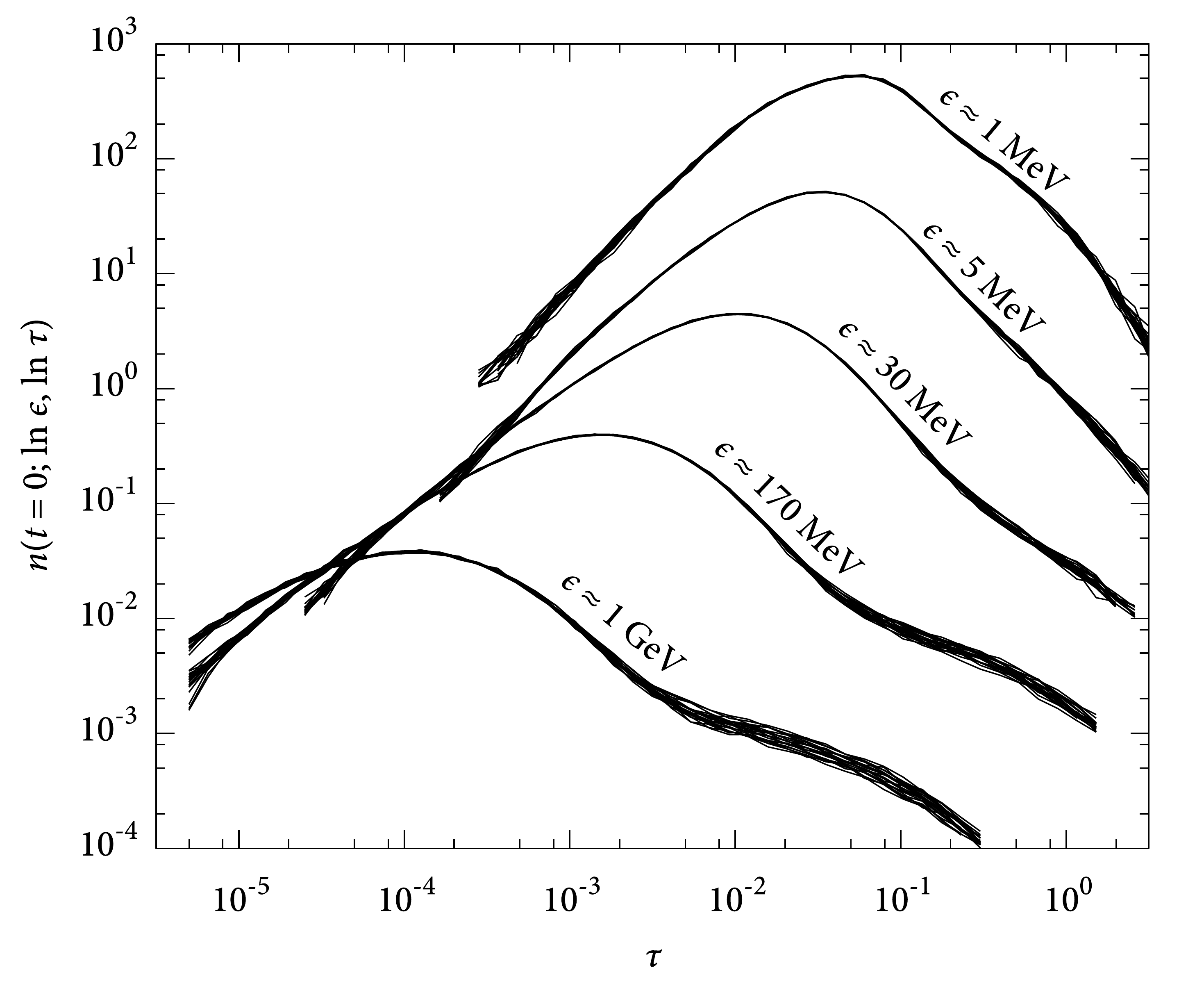}}%
    \caption{Electron distributions~$n(t=0;\ln\epsilon,\ln\tau)$ for different electron energies as a function of delay time for 20 individual showers initiated by $10^{19}$\unit{eV} protons. The curve set for $1$\unit{GeV} is at the actual level; consecutive sets are shifted up by a factor of~10.}
    \label{fig:dist_tau_ind}
\end{figure}
Let us define the delay time~$\Delta t$ of a particle as the time lag with respect to an imaginary particle continuing on the cosmic-ray primary's trajectory with the speed of light in vacuum from the first interaction point. In the distribution of these time lags we must again compensate for differences in Molière radius to obtain a universal description by introducing the variable
\begin{equation}
    \tau\equiv\frac{c\Delta t}{r_\mathrm{M}},
\end{equation}
where~$c$ is the speed of light in vacuum. At sea level, $\tau=1$ corresponds to a time delay of $0.26~\muon$s. The normalised delay time distribution at the shower maximum for different values of~$\epsilon$ is shown in Fig.~\ref{fig:dist_tau_ind} as a function of delay time for 20~individual proton showers. Note the striking resemblance of the time lag distribution to the lateral particle distribution (cf.\ Fig.~\ref{fig:dist_r_ind}). This similarity is a direct result of the non-planar shape of the shower front as discussed in the next section. Therefore, every characteristic in the lateral distribution will have an equivalent in the time lag distribution.

\begin{figure}
    \includeplaatje[width=\figurewidth]{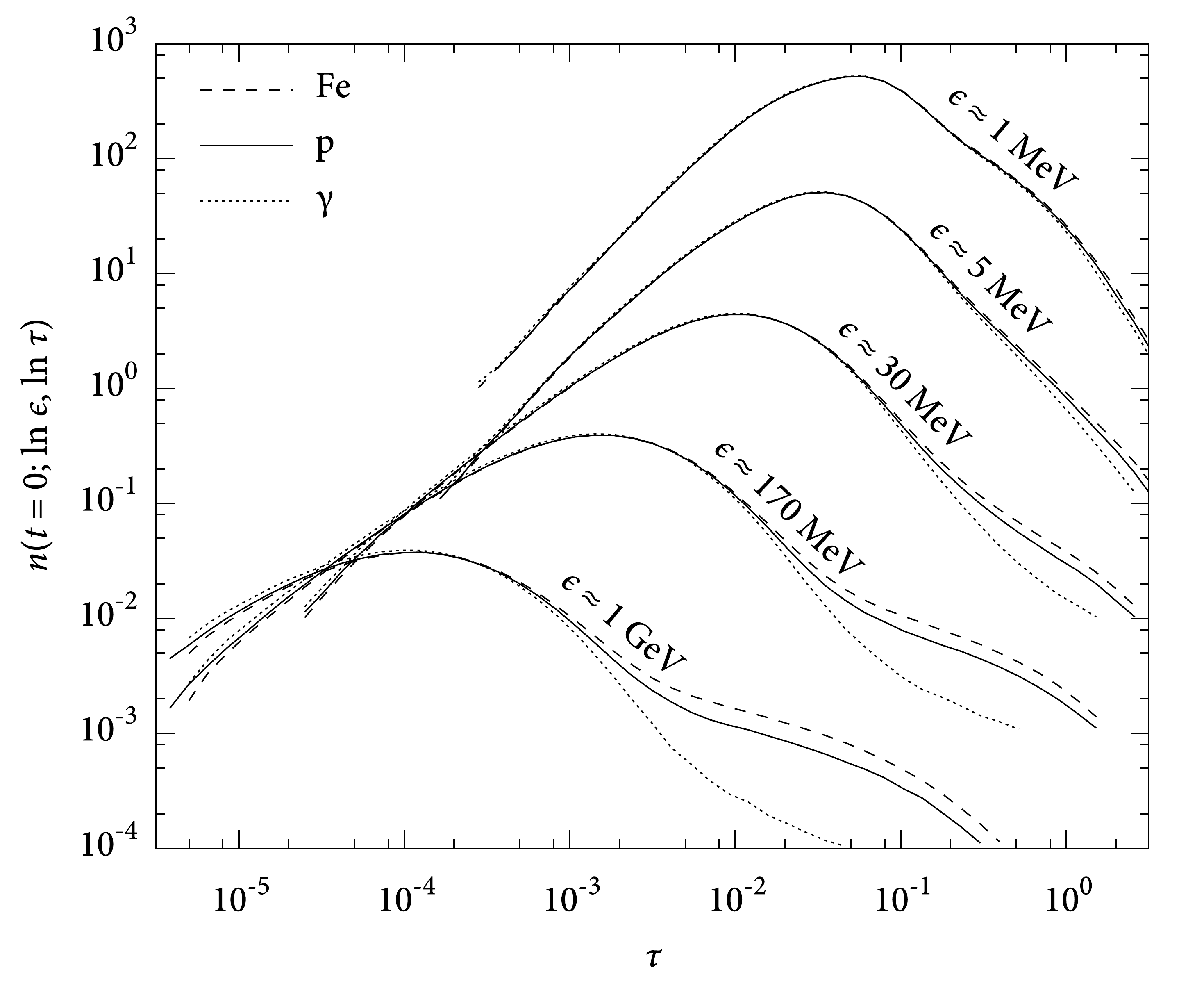}
    \caption{Average distributions~$n(t=0;\ln\epsilon,\ln\tau)$ for different primaries, averaged over 20~showers at $10^{19}$~eV. Consecutive sets are again shifted up by a factor of~10. Note the species-dependent bulge on the right as in Fig.~\ref{fig:dist_r_dep_species}.}
    \label{fig:dist_tau_dep_species}
\end{figure}
The dependencies on primary energy, species, and angle of incidence closely follow those observed in the lateral distributions in every aspect. This includes the behaviour of the second bulge with primary species, as shown in Fig.~\ref{fig:dist_tau_dep_species}. Pion-decay-initiated electrons and positrons are again responsible for the emergence of this peak.

\begin{figure}
    \includeplaatje[width=\figurewidth]{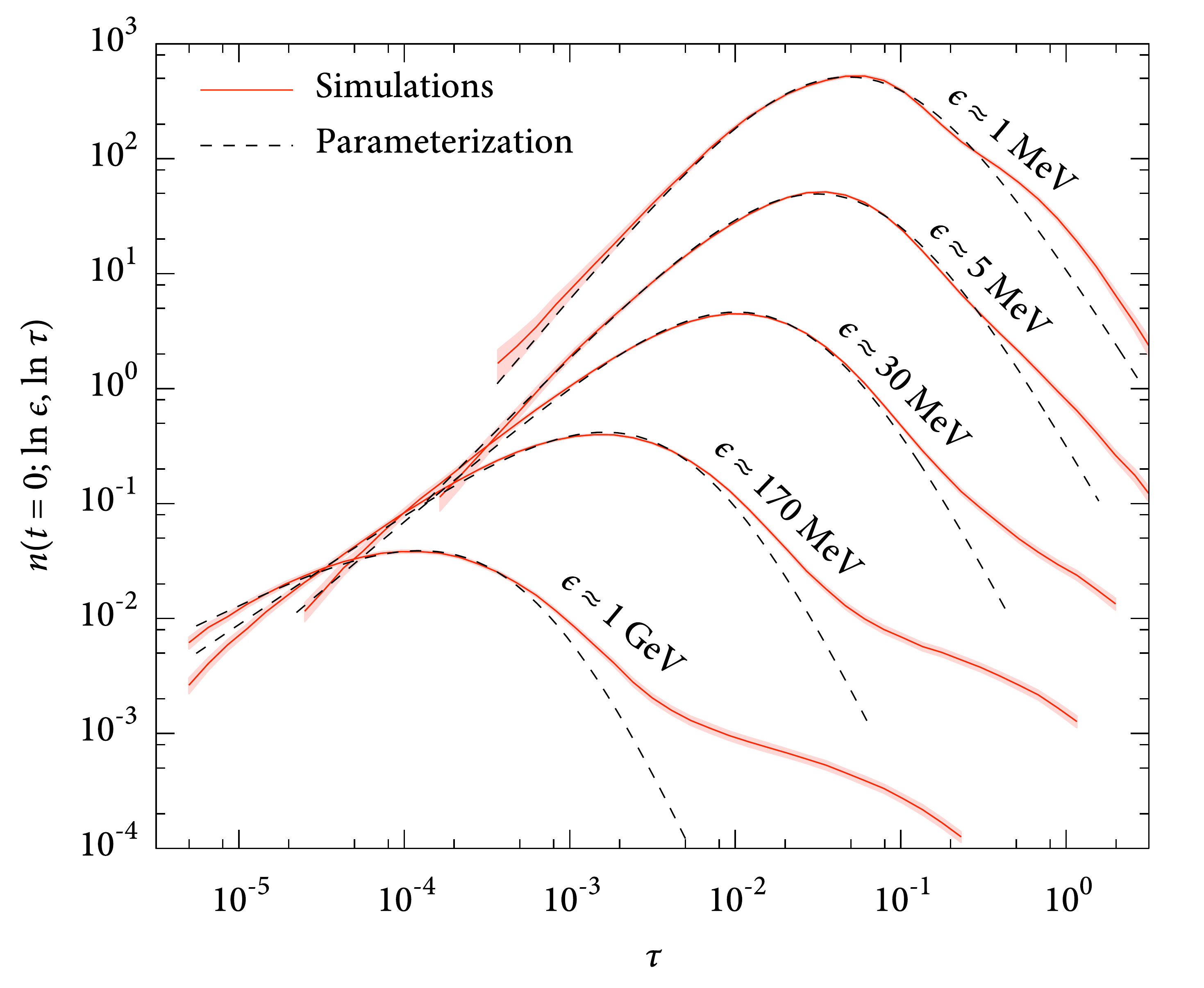}
    \caption{Normalised average electron distributions~$n(t=0;\ln\epsilon,\ln\tau)$ (solid) for 20~proton showers at $10^{19}$\unit{eV} with $3\sigma$~statistical error margins (filled area). For each energy, corresponding parameterizations according to~\eqref{eq:delay} are also drawn (dashed). Consecutive sets are again shifted up by a factor of~10.}
    \label{fig:dist_tau_fit}
\end{figure}

Given the similarity between the lateral and delay time distributions, we use a function of the same form as~\eqref{eq:lateral3d} to parameterize this distribution:
\begin{equation}\label{eq:delay}
    n(t;\ln\epsilon,\ln\tau)=C_3 \tau^{\zeta_0''}(\tau_1+\tau)^{\zeta_1''}.
\end{equation}
Appendix~\ref{app:delay} explains the values of $\tau_i$ and~$\zeta_i''$. Fig.~\ref{fig:dist_tau_fit} compares the parameterization above to the average distribution for proton showers at their maximum. Again, only the main peak was included in defining the fit parameters, causing the resulting parameterized shape to underestimate the number of particles at long delay times.


%
%
\section{Shape of the shower front}
          \label{sec:front}
%
%
The similarity between the lateral and delay time distributions of electrons and positrons as investigated in the previous sections is the result of the spatial extent of an air shower at a given time. It makes sense, therefore, to investigate the physical shape of the shower front by looking at the dependence of the distribution on lateral and delay time simultaneously. In order to keep the analysis practicable, we will abandon energy dependence here in our study.

\begin{figure}
    \centerline{\includeplaatje[width=\figurewidth]{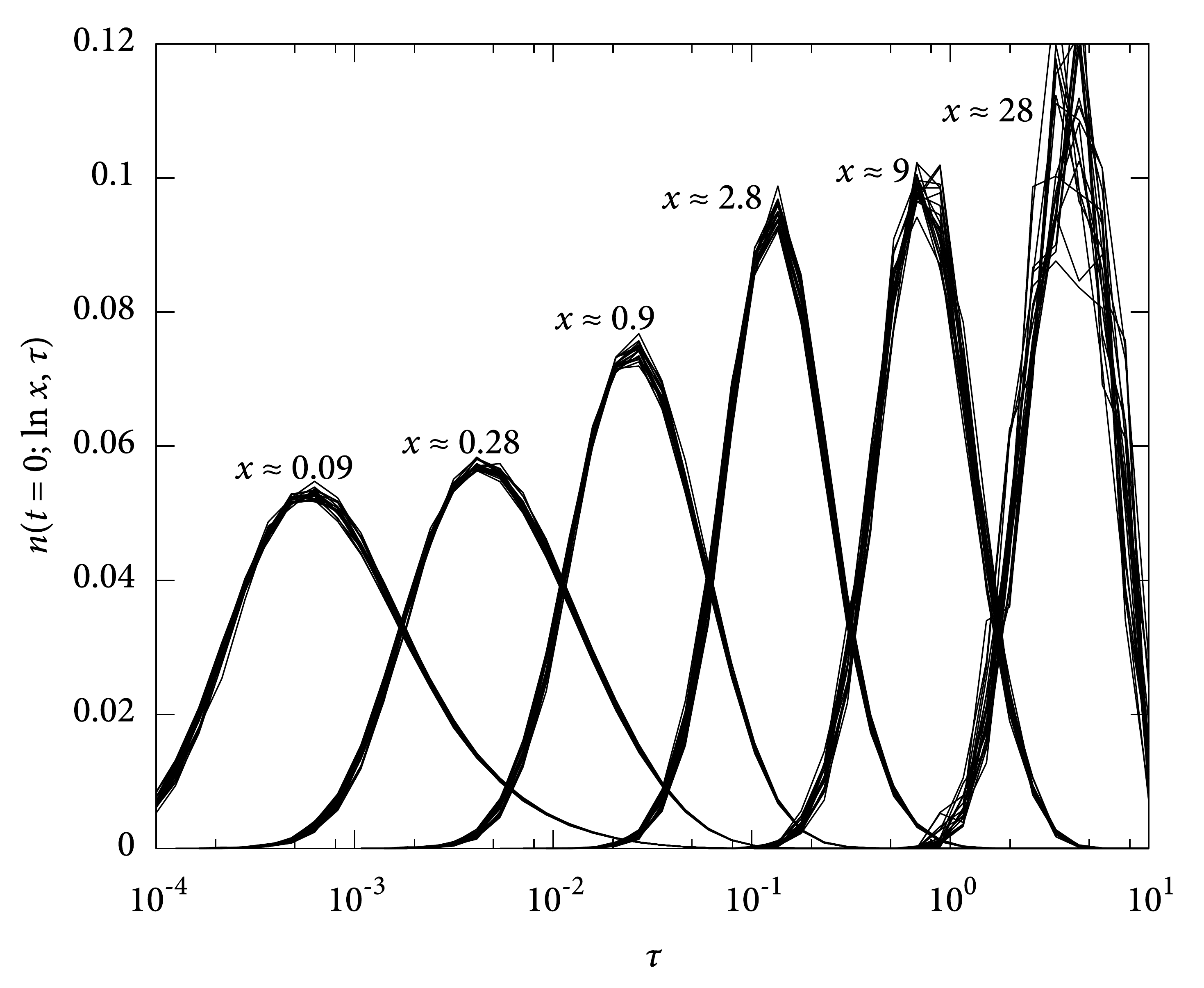}}%
    \caption{Electron distributions~$n(t=0;\ln x,\tau)$ as a function of particle time lag for 20~individual showers initiated by $10^{19}$\unit{eV} protons.}
    \label{fig:front_ind}
\end{figure}
For 20~proton shower simulations at~$10^{19}$\unit{eV}, the shower front shape at the shower maximum is displayed in Fig.~\ref{fig:front_ind} at different distances from the shower core. The distribution shown is~$n(t;\ln x,\tau)$, and each curve is scaled to a similar level for easier comparison of the distributions. Though the low number of particles leads to larger fluctuations of the distributions at high distances, the behaviour clearly does not change significantly for $x>3$.

\begin{figure}
    \centerline{\includeplaatje[width=\figurewidth]{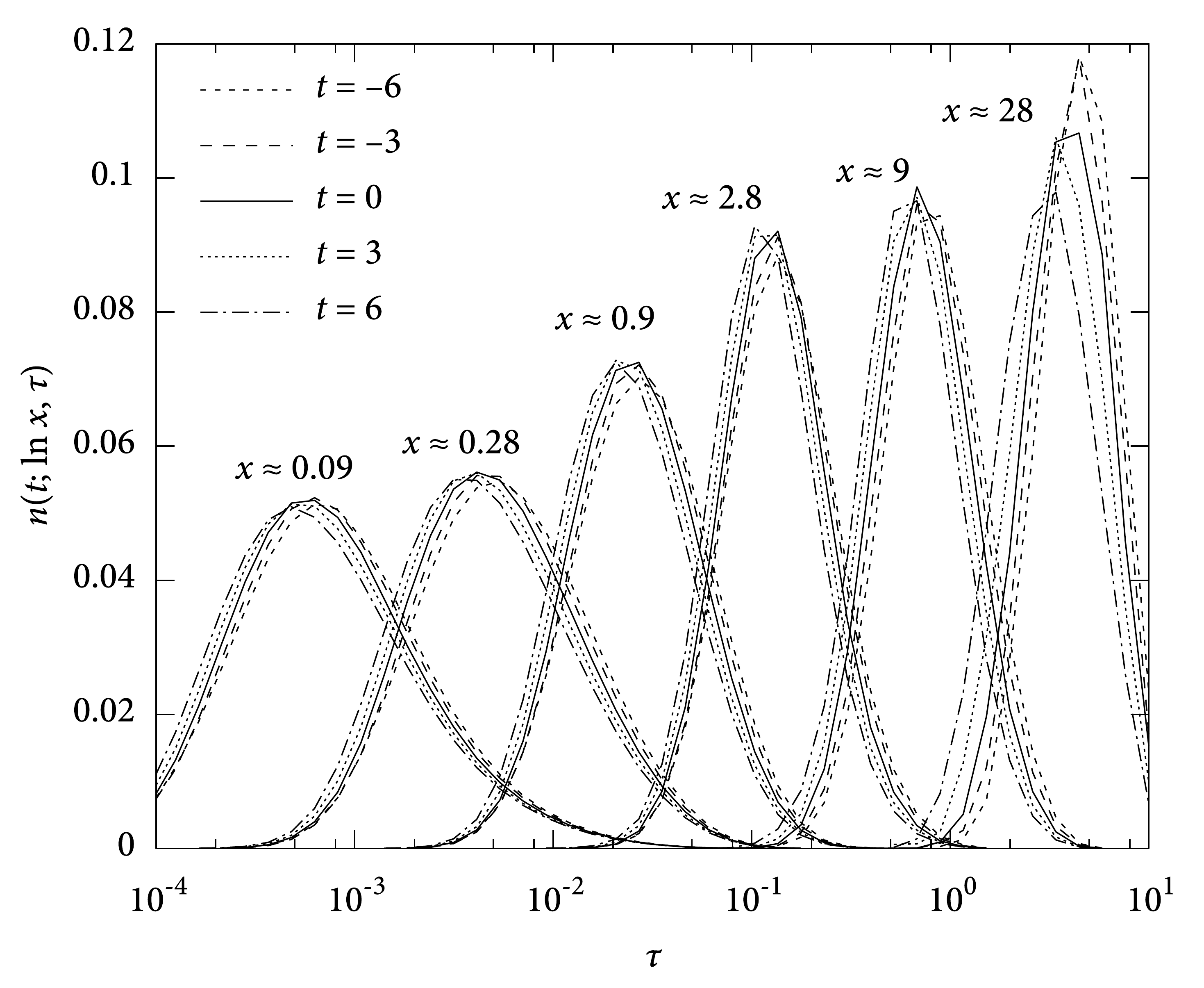}}%
    \caption{Average distributions~$n(t;\ln x,\tau)$ for different evolution stages, averaged over 20~proton-initiated showers at $10^{19}$\unit{eV}.}
    \label{fig:front_dep_age}
\end{figure}
\begin{figure}
    \centerline{\includeplaatje[width=\figurewidth]{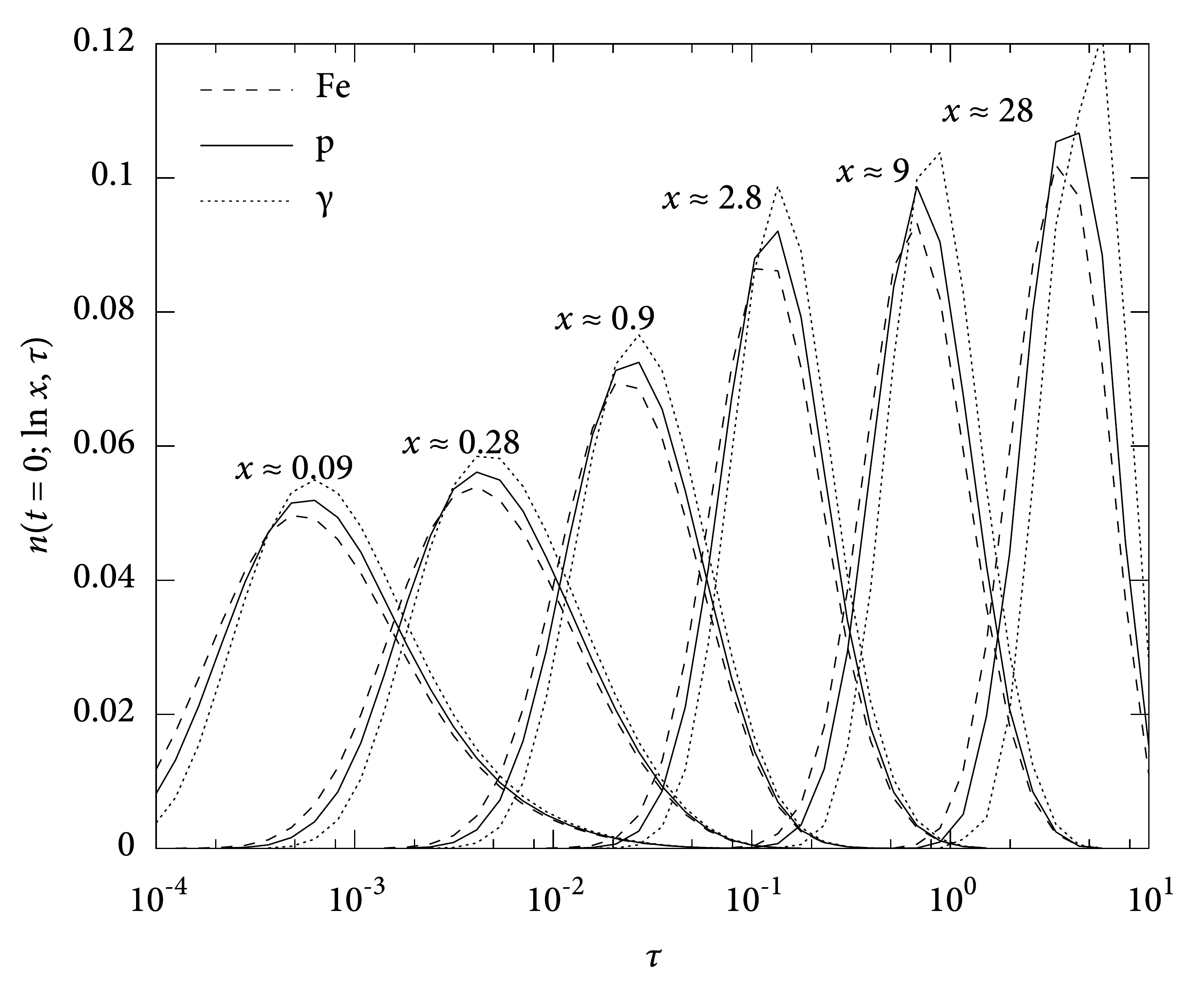}}%
    \caption{Average distributions~$n(t=0;\ln x,\tau)$ for different primary species, averaged over 20~proton-initiated showers at $10^{19}$\unit{eV}.}
    \label{fig:front_dep_species}
\end{figure}
No significant dependence of the shower front shape on incidence angle was found for $x<15$, nor is there any change with primary energy. There are fluctuations with evolution stage, however: the time lag decreases by a constant fraction which depends on the shower stage. As the shower evolves, the entire distribution shifts to the left. This effect, shown in Fig.~\ref{fig:front_dep_age}, can be explained from the increasing spatial structure of the shower with age, not unlike the case of an expanding spherical shell. We shall see further on that the analogy is not entirely legitimate, but the shift does allow one to estimate~$X_\max$ from the arrival times of the particles. We also found a non-negligible dependence of the delay time on primary species, which is comparable in nature to the effect of evolution stage, as shown in Fig.~\ref{fig:front_dep_species}. The dependence of the distribution on both species and evolution stage can be removed almost entirely for distances of $0.03<x<15$ by applying a simple exponential shift in~$\tau$. Additionally, the distributions shown are integrated over energy. Therefore, the shape of the distribution changes when electrons or positrons are considered separately, since their energy distribution is different as well.

The particle distribution at a certain distance from the shower core as a function of arrival time is usually parameterized as a gamma probability density function~\citep{1975:Woidneck,1997:Agnetta}, given by
\begin{equation}\label{eq:nkg}
    n(t;\ln x,\tau) \propto \exp[a_0\ln \tau-a_1\tau].
\end{equation}
We have found that such a parameterization does not follow our simulated distributions very well. Its slope is too gentle at short delay times and too steep at long time lags. Here, we use the better representation
\begin{equation}\label{eq:showerfront}
    n(t;\ln x,\tau) = C_4\exp[a_0\ln \tau'-a_1\ln^2 \tau'],
\end{equation}
which allows for a more gradual slope on the right side of the curve.
The modified time lag~$\tau'$ takes into account the exponential shift mentioned earlier, and is defined as
\begin{equation}\label{eq:tauprime}
    \tau'\equiv\tau\E{-\beta_\mathrm{t} t-\beta_\mathrm{s}},
\end{equation}
where $\beta_\mathrm{t}$ and~$\beta_\mathrm{s}$ are corrections for shower evolution stage and primary species, respectively. The values of the parameters $a_0(x)$, $a_1(x)$, $\beta_\mathrm{t}$, and~$\beta_\mathrm{s}$ are explained in Appendix~\ref{app:showerfront}. The parameter~$\beta_\mathrm{t}$ can be seen as a scale width for the expansion of the shower front as it develops. Note that the integral lateral distribution as parameterized in~\eqref{eq:lateral2d} is needed to obtain actual particle numbers via
\begin{equation}
    N(t;\ln x,\tau)
        = N(t)n(t;\ln x)n(t;\ln x,\tau),
\end{equation}
using the identities in~\eqref{eq:dist_def}.

We may exploit the necessity of the parameter~$\beta_\mathrm{s}$ in our description of the shower front shape to determine the primary species if the value of $X_\max$ is known. To distinguish proton from photon showers in this manner, the required resolution in shower stage is $\delta t<\beta_\mathrm{s}/\beta_\mathrm{t}\simeq0.52$, assuming perfect timing and distance information. This corresponds to an error in $X_\max$ of $19$\unit\gcm. To separate proton from iron showers, the maximum error is reduced to $11$\unit\gcm. Unfortunately, these figures are similar to or smaller than statistical fluctuations in individual showers or systematic uncertainties in the atmospheric density due to weather influences~\cite{2004:Keilhauer,2006:Wilczynska}. This makes it very difficult to take advantage of this intrinsic difference.

\begin{figure}
    \centerline{\includeplaatje[width=\figurewidth]{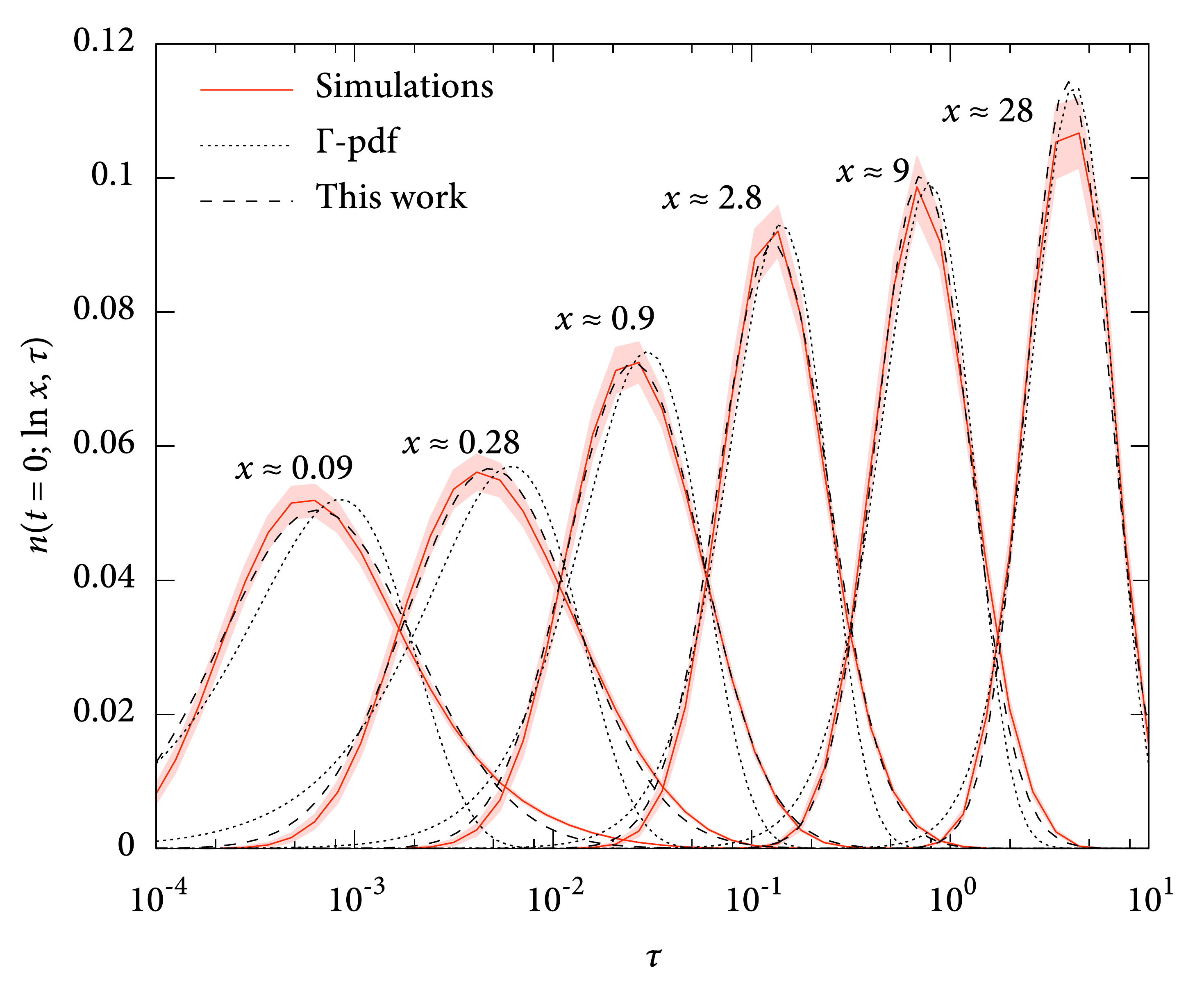}}%
    \caption{Average electron distributions~$n(t=0;\ln x,\tau)$ (solid) for the reference set with $3\sigma$~statistical error margins (filled area). For each distance, corresponding parameterizations according to~\eqref{eq:showerfront} are drawn as well (dashed). Best-fit $\upGamma$-pdf are also plotted (dotted).}
    \label{fig:front_fit}
\end{figure}
An example of the fit of~\eqref{eq:showerfront} at $t=0$ is shown in Fig.~\ref{fig:front_fit}. For distances $x≳0.8$, the fit describes the simulations very accurately. Equivalence is partially lost at small distances, because the shape of the distribution becomes more complicated closer to the shower core. Even there, however, the resulting shape is reasonably accurate down to $x\simeq0.04$. Also plotted are best-fit gamma probability density functions according to~\eqref{eq:nkg} for each distance, which are of lower quality than the parameterization used here, especially close to the core.

\begin{figure}
    \includeplaatje[width=\figurewidth]{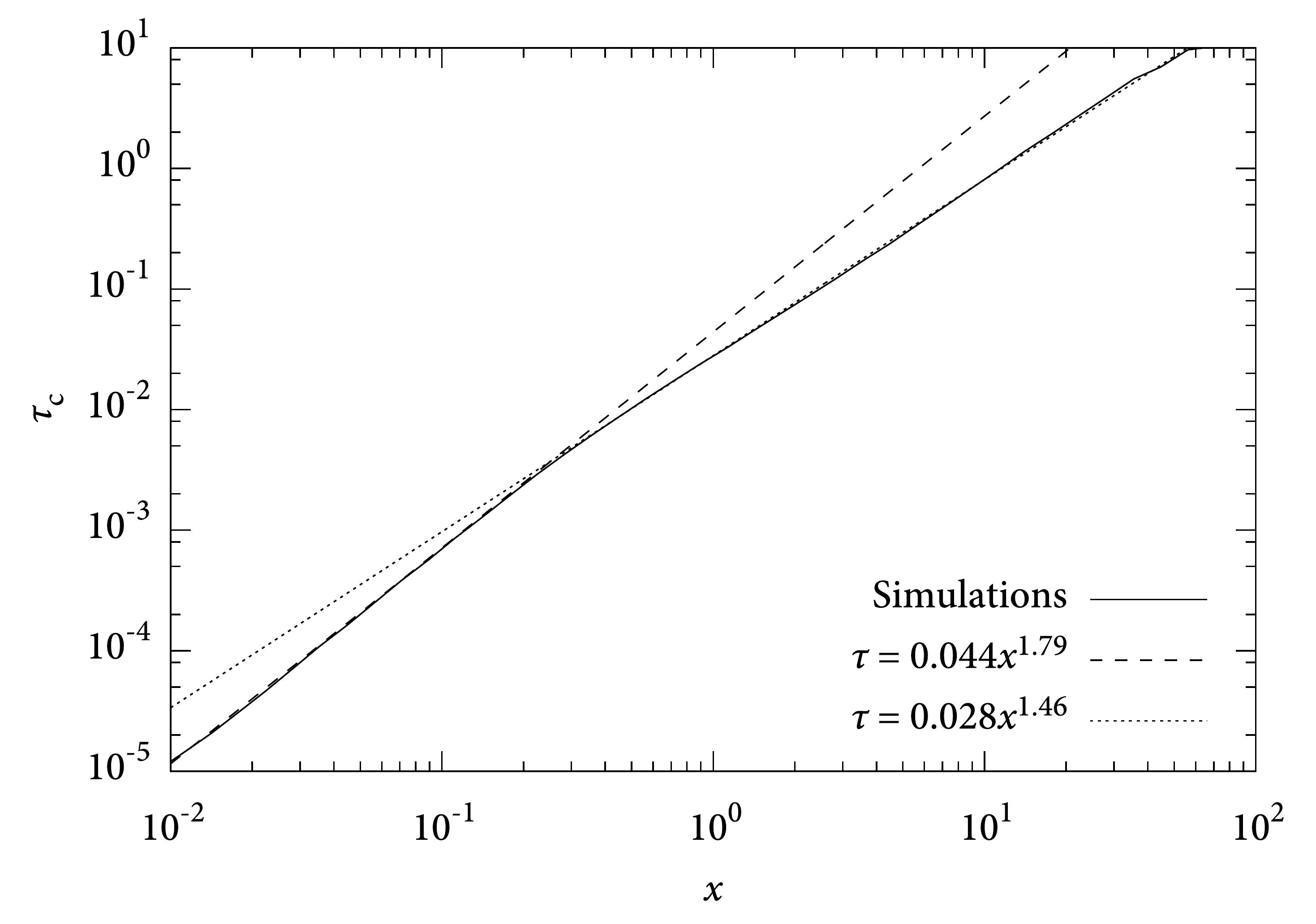}
    \caption{Maximum density~$\tau_\mathrm{c}$ as a function of lateral distance~$x$ at the shower maximum. Also shown are curves for $x<x_0$ (dashed) and $x>x_0$ (dotted) according to the parameterization in~\eqref{eq:showerfrontmax}.}
    \label{fig:front_cutoff}
\end{figure}

For a certain distance from the shower core, we define the time lag~$\tau_\mathrm{c}$ as the time lag where the particle density is at its maximum, corresponding to the peaks of the curves shown earlier in this section. Its value at the shower maximum is shown in Fig.~\ref{fig:front_cutoff} as a function of~$x$ for the reference simulation set. The two straight lines represent fits of the form $\tau_c=Ax^k$ to the part before (dashed) and after the break (dotted) as shown in the plot. The time lag of the maximum particle density can be parameterized as
\begin{equation}\label{eq:showerfrontmax}
\tau_\mathrm{c}=
    \begin{cases}
        (0.044-0.00170t)x^{1.79-0.0056t}&x<x_0;\\
        (0.028-0.00049t)x^{1.46-0.0007t}&x>x_0,\\
   \end{cases}
\end{equation}
where the value for~$x_0$ follows from continuity. One could employ this function to estimate the value of~$X_\max$, though the accuracy attainable in this way is probably much lower than using fluorescence measurements.

In experiments, the shower front is sometimes approximated as a spherical shell~\citep{2004:Dawson}. How do the simulated distributions compare to such a hypothetical shape? Close to the shower core, where $r\ll R$ (with $R\simeq 50$~the supposed curvature radius in Molière units) we expect $k=2$ and $R=A^{-1}$. Going out, the slope should then decrease slowly as~$x$ approaches the presumed curvature radius.

This spherical shape does not correspond to the situation in our simulations. In the innermost region the exponent gives consistently smaller values of~$k\simeq 1.79$. Further out, there is an abrupt transition around $x\simeq0.3$, and the final exponent is $k\simeq 1.45$. 

%
%
\section{Conclusion}
%
%
In this work, we have presented a framework for the accurate description of electron-positron distributions in extensive air showers. To characterize the longitudinal evolution of the air shower, the concept of slant depth relative to the shower maximum is used.

Using the \corsika\ code, we have built a library of simulations of air showers. Analysis of this library shows that, to a large extent, extensive air showers show universal behaviour at very high energy, making the distributions in them dependent on only two parameters: the atmospheric depth~$X_\max$ where the number of particles in the air shower peaks and the total number of particles $N_\max$ present in the shower at this depth. The entire structure of the shower follows directly from these two values.

We have found some exceptions to the universality hypothesis in the spatial distribution of particles. Theoretically, these non-universal features can be employed to distinguish primaries on a shower-to-shower basis. In real experiments, however, this would be a difficult task because the effect either amounts to only a few percent, or its behaviour can be mistaken for variations in shower stage.

To support the simulation of secondary radiation effects from extensive air showers, we have provided two-dimensional parameterizations to describe the electron-positron content in terms of stage vs.\ energy and stage vs.\ lateral distance. We have also supplied three-dimensional representations of the electron content in terms of stage vs.\ energy vs.\ vertical momentum angle, stage vs.\ energy vs.\ horizontal momentum angle, stage vs.\ energy vs.\ lateral distance close to the shower core, and stage vs.\ lateral distance vs.\ arrival time.

Though these parameterizations provide accurate descriptions of the electron-positron distributions in air showers, the authors would like to mention that there are no theoretical grounds for most of the functional representations suggested in this work. Their choice is justified only by the flexibility of the functions to accurately reproduce the simulated distributions as fit functions with a small number of parameters. Additionally, the parameterizations provided are based on simulations with a single interaction model only. Though no significant changes are expected in the general behaviour, the parameters listed will likely change when a different model is employed.

When used together with a longitudinal description for the total number of particles, accurate characterizations of any large air shower in terms of the relevant quantities can be calculated. These may be used for realistic electron-positron distributions without the need for extensive simulations and could be useful in calculations of fluorescence, radio or air Cherenkov signals from very-high-energy cosmic-ray air showers.

%
%
\begin{ack}
The authors wish to express their gratitude to Markus Risse whose help was indispensable in setting up the simulations, and to an anonymous reviewer for useful suggestions to improve the manuscript. One of the authors, R.E., acknowledges fruitful discussions with Paolo Lipari on the subject of shower universality. This work is part of the research programme of the `Stich\-ting voor Fun\-da\-men\-teel Onder\-zoek der Ma\-terie (\textsc{fom})', which is financially supported by the `Neder\-landse Orga\-ni\-sa\-tie voor Weten\-schap\-pe\-lijk Onder\-zoek (\textsc{nwo})'. T.~Huege was supported by grant number VH-NG-413 of the Helmholtz Association.
\end{ack}
%
%
\appendix
\section{Fit parameters}
%
%
This appendix explains in detail the various parameters used in the functional parameterizations throughout this paper. All of these were obtained by performing minimisation sequences using a nonlinear least-squares Marquardt-Levenberg algorithm.
%
%
\subsection{Energy spectrum}\label{app:energy}
\begin{table*}
\caption{Parameter values for the energy spectrum in~\eqref{eq:energy_t} for species of electrons, positrons, and the sum of electrons and positrons.}
\label{tab:energy_par}
\centerline{\begin{tabular}{l@{\quad}l@{\quad}l@{\quad}l@{\quad}l@{\quad}l}
\hline
               & $A_0$ &  $\epsilon_1$  & $\epsilon_2$ & $\gamma_1$ & $\gamma_2$ \\
\hline
Electrons    & $0.485 A_1\exp(0.183t-8.17t^2\e{-4})$ &
                          $3.22-0.0068t$ & $106-1.00t$  & 1          & $1+0.0372t$   \\
Positrons    & $0.516 A_1\exp(0.201t-5.42t^2\e{-4})$ &
                          $4.36-0.0663t$ & $143-0.15t$  & 2          & $1+0.0374t$   \\
Total        &$\hfill A_1\exp(0.191t-6.91t^2\e{-4})$ &
                          $5.64-0.0663t$ & $123-0.70t$  & 1          & $1+0.0374t$   \\
\hline
\end{tabular}}
\end{table*}
The parameters in the energy spectrum distribution function as put forward in~\eqref{eq:energy_t} were chosen to match those advocated in \citet{2006:Nerling}. A good description is obtained with the parameters listed in Table~\ref{tab:energy_par}. The constants in $\epsilon_1$ and $\epsilon_2$ are in~MeV; the constant~$A_0$ is provided here for all three cases to obtain charge excess values; the overall parameter~$A_1$ in the table follows directly from normalisation constraints.

%
%
\subsection{Vertical angular spectrum}\label{app:vertical}

The distribution of the particles' momentum angle away from the shower axis can be parameterized accurately as
$$
    n(t;\ln\epsilon,\Omega)
         = C_0
           \left[\left(\E{b_1}\theta^{\alpha_1}\right)^{-1/\sigma}
             +   \left(\E{b_2}\theta^{\alpha_2}\right)^{-1/\sigma}
           \right]^{-\sigma}.
\eqno{\eqref{eq:vertmom}}$$
For secondary energies $1\unit{MeV}<\epsilon<10$\unit{GeV} and angles up to~$60$º, the curves are described well for $n(t;\ln\epsilon,\Omega)>10^{-4}$ by setting the parameters in the equations above, using nine free parameters, to
\begin{equation}\begin{split}
    b_1       &= -3.73 + 0.92\epsilon^{0.210};\\
    b_2       &= 32.9  - 4.84\ln\epsilon;\\
    \alpha_1  &= -0.399;\\
    \alpha_2  &= -8.36 + 0.440\ln\epsilon.
\end{split}\end{equation}
The constant~$\sigma$ is a parameter describing the smoothness of the transition of the distribution function from the first term of importance near the shower axis to the second term being relevant further away and was set to~$\sigma=3$. The overall factor~$C_0$ follows from the normalisation condition.

%
%
\subsection{Horizontal angular spectrum}\label{app:horizontal}

The horizontal distribution of momentum is given by
$$
    n(t;\ln\epsilon,\phi)
         = C_1[1+\exp(\lambda_0-\lambda_1\phi-\lambda_2\phi^2)],\eqno{\eqref{eq:hormom}}
$$
where optimal agreement is reached in the intervals $1\unit{MeV}<\epsilon<10$\unit{GeV} and $-6<t<9$ by setting
\begin{equation}\begin{split}
    \lambda_0 &= 0.329 - 0.0174 t
               + 0.669\ln\epsilon - 0.0474\ln^2\epsilon;\\
    \lambda_1 &= 8.10\e{-3} + 2.79\e{-3}\ln\epsilon;\\
    \lambda_2 &= 1.10\e{-4} - 1.14\e{-5}\ln\epsilon,
\end{split}\end{equation}
with all energies in~MeV. There were eight free parameters in total in the fit. The value of~$C_1$ follows directly from the normalisation in~\eqref{eq:dist_norm}.

%
%
\subsection{Lateral distribution}\label{app:lateral}

The \textsc{nkg}-like function to describe the primary peak in the lateral distribution is defined as
$$
    n(t;\ln\epsilon,\ln x) =
        C_2' x^{\zeta_0'} (x_1'+x)^{\zeta_1'}.\eqno{\eqref{eq:lateral3d}}
$$
The fit was performed in the interval $1\unit{MeV}<\epsilon<10$\unit{GeV}, with the additional condition that $x<5x_\mathrm{c}$ in order to discard the second, species-dependent peak. Optimal correlation is obtained by using the parameters
\begin{equation}\begin{split}
    x_1'     &= 0.859 - 0.0461\ln^2\epsilon + 0.00428\ln^3\epsilon;\\
    \zeta_t  &= 0.0263t;\\
    \zeta_0' &= \zeta_t + 1.34\\
                &\quad + 0.160\ln\epsilon - 0.0404\ln^2\epsilon + 0.00276\ln^3\epsilon;\\
    \zeta_1' &= \zeta_t - 4.33,
\end{split}\end{equation}
with nine free parameters in total. The value of~$\epsilon$ is always expressed in~MeV.
Again, the value of~$C'_2$ follows directly from normalisation constraints and will not be discussed here.

%
%
\subsection{Delay time distribution}\label{app:delay}

The fit function to describe the primary peak in the delay time distribution is identical to that of the lateral distribution,
$$
    n(t;\ln\epsilon,\ln\tau) =
        C_3 \tau^{\zeta_0''} (\tau_1+\tau)^{\zeta_1''},\eqno{\eqref{eq:delay}}
$$
and was performed at energies $1\unit{MeV}<\epsilon<10$\unit{GeV}, discarding the second peak. Best-fit parameters are
\begin{equation}\begin{split}
    \tau_1    &= \exp[-2.71 + 0.0823\ln\epsilon - 0.114\ln^2\epsilon]\\
    \zeta_0'' &= 1.70 + 0.160t - 0.142\ln\epsilon\\
    \zeta_1'' &= -3.21
\end{split}\end{equation}
with $\epsilon$ in~MeV, using 7~free parameters in total. The constant~$C_3$ again follows from normalisation.

%
%
\subsection{Shape of the shower front}\label{app:showerfront}

The shape of the shower front is parameterized as
$$
    n(t;\ln x,\tau) = C_4\exp[a_0\log \tau'-a_1\log^2 \tau'],\eqno{\eqref{eq:showerfront}},
$$
inspired by the gamma probability distribution, with
$$
    \tau'\equiv\tau\E{-\beta_\mathrm{t} t-\beta_\mathrm{s}}.\eqno{\eqref{eq:tauprime}}
$$
The following parameters give optimal results:
\begin{equation}\begin{split}
    a_0              &= -6.04 + 0.707\log^2 x + 0.210\log^3 x\\
                     &\quad -0.0215\log^4 x -0.00269\log^5 x;\\
    a_1              &= 0.855 + 0.335\log x + 0.0387\log^2 x\\
                     &\quad -0.00662\log^3 x.\\
\end{split}\end{equation}
The value for $\beta_\mathrm{t}$ is fixed at $\beta_\mathrm{t} = 0.20$, while~$\beta_\mathrm{s}$ depends on the primary species:
\begin{align}
    \beta_\mathrm{s} &=     -0.062    &&\text{for iron nuclei;}\nonumber\\
    \beta_\mathrm{s} &\equiv 0         &&\text{for protons;}\\
    \beta_\mathrm{s} &=      0.103     &&\text{for photons.}\nonumber
\end{align}
These parameters are valid for distances of $0.4<x<10^2$ and $10^{-4}<\tau'<10$. 

%
%
\bibliographystyle{unsorted}
\enlargethispage{2.5ex}
\begin{small}
\bibliography{/Users/svenlafe/Documents/Werk/literatuur/svenlafe}
\end{small}
\end{document}